\documentclass[sigconf,natbib=true,anonymous=false]{acmart}
\usepackage{amsmath,amsfonts}  
\usepackage{multirow} 
\usepackage{subfigure} 
\usepackage{algorithm}
\usepackage{algorithmic}
\usepackage{enumitem}

\AtBeginDocument{%
  \providecommand\BibTeX{{%
    \normalfont B\kern-0.5em{\scshape i\kern-0.25em b}\kern-0.8em\TeX}}}

\copyrightyear{2023}
\acmYear{2023}
\copyrightyear{2023}
\acmYear{2023}
\setcopyright{acmlicensed}\acmConference[SIGIR '23]{Proceedings of the 46th International ACM SIGIR Conference on Research and Development in Information Retrieval}{July 23--27, 2023}{Taipei, Taiwan}
\acmBooktitle{Proceedings of the 46th International ACM SIGIR Conference on Research and Development in Information Retrieval (SIGIR '23), July 23--27, 2023, Taipei, Taiwan}
\acmPrice{15.00}
\acmDOI{10.1145/3539618.3591727}
\acmISBN{978-1-4503-9408-6/23/07}

\begin{document}

\title{Meta-optimized Contrastive Learning for Sequential Recommendation}


\author{Xiuyuan Qin}
\authornote{These authors contributed equally to this work.}
\affiliation{
  \city{Soochow University}
  \country{China}}
\email{20215227016@stu.suda.edu.cn}

\author{Huanhuan Yuan}
\authornotemark[1]
\affiliation{
  \city{Soochow University}
  \country{China}}
\email{hhyuan@stu.suda.edu.cn}

\author{Pengpeng Zhao}
\authornote{Corresponding author.}
\affiliation{
  \city{Soochow University}
  \country{China}}
\email{ppzhao@suda.edu.cn}

\author{Junhua Fang}
\affiliation{
  \city{Soochow University}
  \country{China}}
\email{jhfang@suda.edu.cn}

\author{Fuzhen Zhuang}
\affiliation{
  \institution{Institute of Artificial Intelligence \& SKLSDE, School of Computer Science}
  \city{Beihang University}
  \country{China}}
\email{zhuangfuzhen@buaa.edu.cn}

\author{Guanfeng Liu}
\affiliation{
  \city{Macquarie University}
  \country{Australia}}
\email{guanfeng.liu@mq.edu.au}


\author{Victor Sheng}
\affiliation{
  \city{Texas Tech University}
  \country{United States}}
\email{victor.sheng@ttu.edu}







\renewcommand{\shortauthors}{Xiuyuan Qin et al.}

\begin{abstract}
Contrastive Learning (CL) performances as a rising approach to address the challenge of sparse and noisy recommendation data. Although having achieved promising results, most existing CL methods only perform either hand-crafted data or model augmentation for generating contrastive pairs to find a proper augmentation operation for different datasets, which makes the model hard to generalize. Additionally, since insufficient input data may lead the encoder to learn collapsed embeddings, these CL methods expect a relatively large number of training data (e.g., large batch size or memory bank) to contrast. However, not all contrastive pairs are always informative and discriminative enough for the training processing. 
Therefore, a more general CL-based recommendation model called Meta-optimized Contrastive Learning for sequential Recommendation (MCLRec) is proposed in this work. By applying both data augmentation and learnable model augmentation operations, this work innovates the standard CL framework by contrasting data and model augmented views for adaptively capturing the informative features hidden in stochastic data augmentation. Moreover, MCLRec utilizes a meta-learning manner to guide the updating of the model augmenters, which helps to improve the quality of contrastive pairs without enlarging the amount of input data. Finally, a contrastive regularization term is considered to encourage the augmentation model to generate more informative augmented views and avoid too similar contrastive pairs within the meta updating. The experimental results on commonly used datasets validate the effectiveness of MCLRec\footnote{Our code is available at \url{https://github.com/QinHsiu/MCLRec}.
}.
\end{abstract}

\maketitle
\section{Introduction}
Sequential Recommendation (SR) models are designed to predict a user's next interacted item based on his/her historical interaction sequence~\cite{SRS,SSL}. Compared with other types of recommender systems, SR could accurately characterize users' dynamic interest in the long- and short-term, and capture the sequential pattern hidden in the users' behaviors. Although many of them, such as GRU4Rec~\cite{GRU4Rec}, SASRec~\cite{SASRec}, and BERT4Rec~\cite{BERT4Rec}, have achieved immense performance improvements, they are limited by the sparse and noisy data. Recently, Contrastive Learning (CL) based recommender systems, which leverage the information from different views to boost the effectiveness of learned representations, are introduced to cope with these problems~\cite{SGL}.

We recap the basic idea of Contrastive Learning (CL), which is to learn an expressiveness embedding by maximizing agreement between the augmented views of the same sequence and pushing away the views of different sequences. Thus, the choice of augmentation operations becomes one of the most crucial problems for every CL recommendation model. Particularly, according to the different augmentation operations, CL-based recommender systems can be divided into three categories. The first category~\cite{CL4SRec,CoSeRec,ICLRec} generates different views of the same sequential data by manually choosing random `mask', `crop', or `reorder' operations on the data level. And the second one~\cite{DuoRec} produces contrastive pairs by `dropout' on the model level. And the third one~\cite{SRMA} combines three model augmentation methods (i.e., `neural mask', `layer drop', and `encoder complement') with data augmentation for constructing view pairs. Most of them are designed as auxiliary tasks to help the primary task of improving the recommendation accuracy.

Despite the recent advances, almost all current CL-based models produce contrastive pairs by manually identifying which random augmentation operations are conducted on either the data level or model level. Nevertheless, the need for augmentation operations in practice for different datasets always be diverse and evolving. Even though sequence augmentation methods utilizing random sequence or model perturbations (including `crop', and `mask' operations) have been widely used and shown great superiority, only relying on such unlearnable operations often requires domain expertise and hand-crafted design, which may not be enough to search for a suitable augmentation operation in such a setting. 

Furthermore, self-supervised contrastive learning does not require labeled data, but insufficient input data may lead the encoder to learn collapsed embeddings~\cite{Barlow, BYOL}. Conventionally, contrastive methods enlarge the batch (or memory bank) and increase the number of augmented views to promote the performance of models for better representations, but many contrastive pairs maybe not be too informative to guide the training, i.e., the representations of positive pairs are pretty close, and negative pairs are already very apart in the latent space~\cite{MetAug}. Such pairs may have few contributions to the optimization and lead to contrastive methods further to pursue the large numerous input data to collect informative ones. Moreover, simply enlarging the batch size will highly promote the cost and reduce training efficiency.

In this paper, we propose a general sequential recommendation model with a meta-learning algorithm, which we call Meta-optimized Contrastive Learning for sequential Recommendation (MCLRec). Firstly, auxiliary contrastive learning is chosen to complement the primary task in both the data and model perspectives. A learnable model augmentation method is combined with data augmentation methods in MCLRec for extracting more expressive features. In this way, model augmented views can serve as additional contrastive pairs and be contrasted with data augmented views during training. Additionally, the parameters of the model augmenters could adaptively adjust to different datasets. Secondly, we leverage a meta manner to update the parameters of the augmentation model according to the performance of the encoder. By using such a learning paradigm, the augmentation model could learn discriminative augmented views based on a relatively restricted amount of interactions (e.g., small batch size).
Finally, a contrastive regularization term is considered in MCLRec by injecting a margin between the similarities of similar pairs for avoiding feature collapse and generating more informative and discriminative features. In a short, the major contributions of MCLRec are as follows: 
\begin{itemize}[leftmargin=*] 
\item A learnable contrastive learning method MCLRec is proposed for sequential recommendation. MCLRec extracts additional helpful information from the existing positive and negative samples (generated by data augmentation) by combining data augmentation and learnable model augmentation.
\item A meta-optimized manner is leveraged in the proposed model MCLRec to guide the training of learnable model augmenters and help learn more discriminative features for recommendation.
\item Extensive experiments on different public benchmark datasets demonstrate that MCLRec can significantly outperform the state-of-the-art sequential methods.
\end{itemize}
\section{Preliminaries}
\subsection{Problem Definition}
Sequential Recommendation (SR) is to recommend the next item that the user will interact with based on his/her historical interaction data. Assuming that user sets and item sets are $\mathcal{U}$ and $\mathcal{I}$ respectively, user $u \in \mathcal{U}$ has a sequence of interacted items $S^{u}=\{i^{u}_{1},..., i^{u}_{|S^{u}|} \}$ and $i^{u}_{k} \in \mathcal{I} (1 \leq k \leq |S^{u}|)$ represents an interacted item at position $k$ of user $u$ within the sequence, where $|S^{u}|$ denotes the sequence length. Given the historical interactions $S^{u}$, the goal of SR is to recommend an item from the set of items $\mathcal{I}$ that the user $u$ may interact with at the $|S^{u}|+1$ step:
 \begin{equation}
 \arg\max_{i \in \mathcal{I}} P(i^{u}_{|S^{u}+1|}=i|S^{u})
     \label{eq0}
 \end{equation}
\subsection{Sequential Recommendation Model}
The backbone SR model used in our model contains three parts, (1) embedding layer, (2) representation learning layer, and (3) next item prediction layer.
\subsubsection{\textbf{Embedding Layer}.} Firstly, the whole item sets $\mathcal{I}$ are embedded into the same space~\cite{SASRec,BERT4Rec} and generate the item embedding matrix $\mathbf{M} \in \mathbb{R}^{|\mathcal{I}| \times d}$. Given the input sequence $S^{u}$, the embedding of the sequence $S^{u}$ is initialized to $\mathbf{e}^{u} \in \mathbb{R}^{n \times d}$ and $\mathbf{e}^{u}=\{\mathbf{m}_{s_{1}}+\mathbf{p}_{1},\mathbf{m}_{s_{2}}+\mathbf{p}_{2},...,\mathbf{m}_{s_{n}}+\mathbf{p}_{n}\}$, where $\mathbf{m}_{s_{k}} \in \mathbb{R}^{d}$ represents the item's embedding at the position $k$ in the sequence, $\mathbf{p}_{k} \in \mathbb{R}^{d}$ represents the position embedding in the sequence and $n$ represents the length of the sequence.
\subsubsection{\textbf{Representation Learning Layer}.} Given the sequence embedding $\mathbf{e}^{u}$, a deep neural network model (e.g., SASRec~\cite{SASRec}) represented as $f_{\theta}(\cdot)$ is utilized to learn the representation of the sequence. Where $\theta$ represents the parameters of the sequential model. The output representation $\mathbf{H}^{u} \in \mathbb{R}^{n \times d}$ is calculated as:
\begin{equation}
\mathbf{H}^{u}=f_{\theta}(\mathbf{e}^{u}).
    \label{eq1}
\end{equation}
The last vector $\mathbf{h}^{u}_{n} \in \mathbb{R}^{d}$ in $\mathbf{H}^{u}=[\mathbf{h}^{u}_{0},\mathbf{h}^{u}_{1},...,\mathbf{h}^{u}_{n}]$ is chosen as the representation of the sequence~\cite{DuoRec}.
\subsubsection{\textbf{Next Item Prediction Layer}.} Finally, the probability of each item $\hat{\mathbf{y}} =\mathrm{softmax}(\mathbf{h}^{u}_{n}\mathbf{M}^\top)$, where $\hat{\mathbf{y}} \in \mathbb{R}^{|\mathcal{I}|}$.
A cross-entropy loss is optimized to maximize the probability of correct prediction:
\begin{equation}
        \mathcal{L}_{rec}=-1*\hat{\mathbf{y}}[g]+\log(\sum_{i\in \mathcal{I}}\exp(\hat{\mathbf{y}}[i]))),
    \label{eq2}
\end{equation}
where $g \in \mathcal{I}$ represents the ground-truth of user $u$.
\begin{figure*}[t]
\centering 
\includegraphics[width=\linewidth]{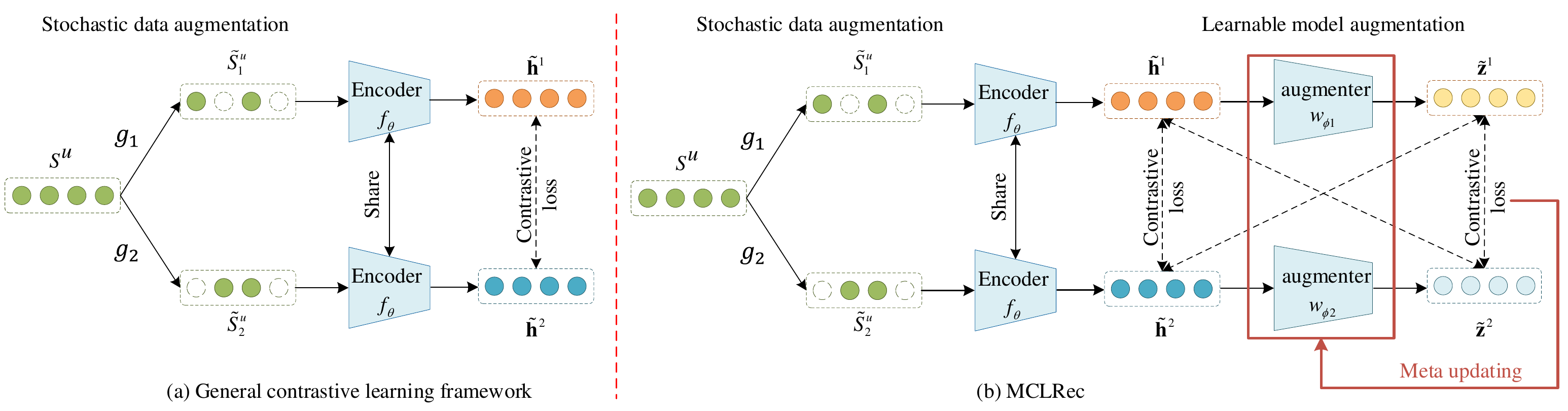}
    \caption{General contrastive learning framework (a) $vs.$ MCLRec (b). In general contrastive learning, the applied data augmentation operations are randomly chosen and the generated augmented views are directly contrasted with each other. In MCLRec, we utilize the learnable augmenters to generate two more model augmented views for contrastive learning and leverage the contrastive loss to guide the training of these augmenters in a meta-optimized manner.}
    \label{model}
\end{figure*}
\section{Methodology}
As shown in Figure~\ref{model}(a), a general contrastive learning framework commonly consists of a stochastic data augmentation module, a user representation encoder, and a contrastive loss function~\cite{CL4SRec}. Different from this general CL paradigm that only relies on data augmentation operation, MCLRec further leverages two learnable augmenters to find the suitable augmentation operation adaptively. The whole framework of the MCLRec is depicted in Figure~\ref{model}(b). Especially, MCLRec consists of three main parts, (1) augmentation module, (2) meta-learning training strategy, and (3) contrastive regularization. All these three modules would be elaborated on in the following subsections.
\subsection{Augmentation Module}
 Our augmentation module mainly contains two parts, a stochastic data augmentation module, and a learnable model augmentation module. The former is to generate two different augmented sequences from the same sequence, and the latter is to capture more informative features according to these augmented sequences.
 \subsubsection{\textbf{Stochastic Data Augmentation Module}.} As shown in Figure~\ref{model}(b), the first half of our model is the same as the general contrastive learning framework~\cite{CL4SRec,CoSeRec}. The stochastic data augmentation operations in MCLRec could be any classical augmentation e.g., `mask', `crop', or `reorder' operations, to create two positive views of a sequence. Given a sequence $S^{u}$, and a pre-defined data augmentation function set $\mathcal{G}$, we denote the generation of two positive views as follows:
\begin{equation}
\Tilde{S}^{u}_{1}=g_{1}(S^{u}), \Tilde{S}^{u}_{2}=g_{2}(S^{u}), \,s.t.\, g_{1},g_{2}\sim\mathcal{G},
    \label{eq3}
\end{equation}
where $g_1$ and $g_2$ represent the data augmentation functions sampled from $\mathcal{G}$, $\Tilde{S}^{u}_{1}$ and $\Tilde{S}^{u}_{2}$ denote the different augmented sequences. Taking $\Tilde{S}^{u}_{1}$ and $\Tilde{S}^{u}_{2}$ as inputs, the data augmentation views $\Tilde{\mathbf{h}}^{1}$ and $\Tilde{\mathbf{h}}^{2}$ are generated according to Eq.~(\ref{eq1}).

\subsubsection{\textbf{Learnable Model Augmentation Module}.}  
General contrastive methods only rely on data augmentations.
More recent emerging contrastive methods leverage different dropouts to generate different augmentation models for constructing contrastive loss~\cite{DuoRec}, which provides a new way to produce augmentation views and inspires us to take advantage of both data and model augmentation.
However, no matter the previous data or model augmentation, their inflexible and random augmentations are hard to generalize in practice. That makes the adaptive and learnable augmentation needed for the CL framework. 
Hence, we propose to use two learnable augmenters to capture the informative features hidden in the stochastic data augmented views. 

As shown in Figure~\ref{model}(b), $\Tilde{\mathbf{h}}^{1}$ and $\Tilde{\mathbf{h}}^{2}$ are fed into the augmentation model $w_{\phi1}(\cdot)$ and $w_{\phi2}(\cdot)$, respectively. The model augmentation views $\Tilde{\mathbf{z}}^{1}$ and $\Tilde{\mathbf{z}}^{2}$ are calculated as:
\begin{equation}
\Tilde{\mathbf{z}}^{1}=w_{\phi1}(\Tilde{\mathbf{h}}^{1}),\Tilde{\mathbf{z}}^{2}=w_{\phi2}(\Tilde{\mathbf{h}}^{2}),
    \label{eq4}
\end{equation}
where $\phi1$ and $\phi2$ represent the parameters of two augmenters, which enable the augmentation operation to be learned end to end and adaptively find optimal augmenters for different datasets. The newly generated $\Tilde{\mathbf{z}}^{1}$ and $\Tilde{\mathbf{z}}^{2}$ could act as contrastive pairs to produce more augmentation views without enlarging batch size. Due to the powerful capability of approximating function, the simple Multi-Layer Perceptron (MLP)~\cite{mlp} is chosen as the augmentation model of MCLRec. We leave other neural network models such as self-attention for future work studies. 

\subsection{Meta-Learning Training Strategy} 
After the introduction of learnable model augmenters, there are two modules that with parameters need to be updated, each with its own objective (i.e., multi-task learning for the encoder and contrastive task for the augmenters). Since there is possibly a gap between these two objectives, directly updating their parameters using joint learning may lead to suboptimal solutions~\cite{multi}.
Therefore, we follow~\cite{MAML,Meta} to perform a meta-learning strategy to guide the training of two augmenters, which is beneficial for the model to mine discriminative augmentation views from the sequence. The whole training process can be concluded in two stages.

In the first stage, we contrast all four augmented views (i.e., $\Tilde{\mathbf{h}}^{1}$, $\Tilde{\mathbf{h}}^{2}$, $\Tilde{\mathbf{z}}^{1}$ and $\Tilde{\mathbf{z}}^{2}$) in different ways and unite the recommendation loss to update the parameters of encoder $f_{\theta}(\cdot)$. 
\begin{table}[t]
  \centering
    \caption{Comparison with other contrastive learning models.}
    \renewcommand{\arraystretch}{1.2}
  \resizebox{1.0\linewidth}{!}{
    \begin{tabular}{c|cc|c|c|c|c|c|c|c}
    \hline
    \multicolumn{3}{c|}{Augmentation Type} & CL4SRec & CoSeRec&LMA4Rec & ICLRec & DuoRec&SRMA & Ours \\
    \hline
    \multicolumn{1}{c|}{\multirow{2}[1]{*}{Stochastic}} & \multicolumn{2}{c|}{Data Level} & $\checkmark$     & $\checkmark$     & $\checkmark$ &  $\checkmark$  & $\times$ & $\checkmark$    & $\checkmark$ \\
    \cline{2-10} & \multicolumn{2}{c|}{Model Level} & $\times$     & $\times$  &$\times$   & $\times$     & $\checkmark$ &$\checkmark$     & $\times$ \\
    \hline
    Learnable & \multicolumn{2}{c|}{Model Level} & $\times$     & $\times$   &$\checkmark$  & $\times$     & $\times$ &$\times$     & $\checkmark$ \\
    \hline
    \end{tabular}}%
  \label{tab:contrast}%
\end{table}%
In the second stage, we utilize the learned encoder $f_{\theta'}(\cdot)$ to re-encode the sequence and use the contrastive loss related to the augmenters to optimized update $w_{\phi1}(\cdot)$ and $w_{\phi2}(\cdot)$. Concretely, $f_{\theta}(\cdot)$, $w_{\phi1}(\cdot)$ and $w_{\phi2}(\cdot)$ are iteratively trained until convergence. 
Especially, in the first stage, we randomly initialize the parameters of encoder $f_{\theta}(\cdot)$ and two augmenters $w_{\phi1}(\cdot)$ and $w_{\phi2}(\cdot)$. After getting all four augmented views, we calculate the recommendation loss by Eq.~(\ref{eq2}) and joint contrastive losses to update the encoder $f_{\theta}(\cdot)$ by back-propagation, which can be calculated as:
\begin{equation}
\mathcal{L}_{0}=\mathcal{L}_{rec}+\lambda\mathcal{L}_{cl1}+\beta\mathcal{L}_{cl2},
    \label{eq5}
\end{equation}
where $\lambda$ and $\beta$ are hyper-parameters that need to be tuned. $\mathcal{L}_{cl1}$ and $\mathcal{L}_{cl2}$ denote the two kinds of contrastive losses. The first kind of contrastive loss acts as the same role as infoNCE loss of general contrastive learning~\cite{SimCLR,Moco}, called $\mathcal{L}_{cl1}$, which depends only on data augmented view $\Tilde{\mathbf{h}}^{1}$ and $\Tilde{\mathbf{h}}^{2}$. It can be formulated as:
\begin{equation}
\mathcal{L}_{cl1}=\mathcal{L}_{con}(\Tilde{\mathbf{h}}^{1},\Tilde{\mathbf{h}}^{2}),
\label{eq6}
\end{equation}
and
\begin{equation}
\begin{split}
\mathcal{L}_{con}(\mathbf{x}^{1},\mathbf{x}^{2})=-\log\frac{e^{s(\mathbf{x}^{1}, \mathbf{x}^{2})}}{e^{s(\mathbf{x}^{1}, \mathbf{x}^{2})}+\underset{\mathbf{x} \in neg}{\sum}e^{s(\mathbf{x}^{1},\mathbf{x})}}\\
-\log\frac{e^{s(\mathbf{x}^{2}, \mathbf{x}^{1})}}{e^{s(\mathbf{x}^{2}, \mathbf{x}^{1})}+\underset{\mathbf{x} \in neg}{\sum}e^{s(\mathbf{x}^{2},\mathbf{x})}},
\end{split}
    \label{eq7}
\end{equation}
where $(\mathbf{x}^{1}, \mathbf{x}^{2})$ represents a pair of positive sample's embedding, $s(\cdot)$ represents inner product and $neg$ indicates the negative sample embedding set. The positive pairs obtained from the same sequence and other 2($|\mathbf{B}|$-1) views within the same batch are treated as negative samples, where $|\mathbf{B}|$ denotes the batch size. 
The second kind of contrastive learning loss called $\mathcal{L}_{cl2}$, is generated from both data and model augmentation views. It can be calculated as:
\begin{equation}
  \mathcal{L}_{cl2}=\mathcal{L}_{con}(\Tilde{\mathbf{z}}^{1},\Tilde{\mathbf{z}}^{2})+\mathcal{L}_{con}(\Tilde{\mathbf{h}}^{1},\Tilde{\mathbf{z}}^{2})+\mathcal{L}_{con}(\Tilde{\mathbf{h}}^{2},\Tilde{\mathbf{z}}^{1}).  
    \label{eq8}
\end{equation}

In the second stage, we fix the parameters of encoder $f_{\theta}(\cdot)$ and optimize $w_{\phi1}(\cdot)$ and $w_{\phi2}(\cdot)$ with respect to the performance of the encoder.
Denote $\theta'$ is the learned parameters by back-propagation at the first stage, we use the learned encoder
$f_{\theta'}(\cdot)$ to re-encode the augmented sequence by Eq.~(\ref{eq1}), recompute $\mathcal{L}_{cl2}$ by Eq.~(\ref{eq8}), and then leverage back-propagation to update the augmenters. The loss is calculated as follows:
\begin{equation}
\mathcal{L}_{1}=\mathcal{L}_{cl2}.
    \label{eq9}
\end{equation}
Then we get learned augmenters $w_{\phi1'}(\cdot)$ and $w_{\phi2'}(\cdot)$, where $\phi1'$ and $\phi2'$ represent the learned parameters by back-propagation at second stage. 
\begin{algorithm}[t] 
\caption{The MCLRec Algorithm} 
\label{alg}
\begin{itemize}[leftmargin=*]
\item[]{\bf{Input:}} Training dataset $\{S_{u}\}_{u=1}^{|U|}$, learning rate $l$ and $l'$, hyper-\\parameters $\lambda, \beta, \gamma$;
\item[]{\bf{Initialize:}} $\theta$ for encoder $f_{\theta}(\cdot)$, $\phi1$ for augmenter $w_{\phi1}(\cdot)$ and $\phi2$ for augmenter $w_{\phi2}(\cdot)$;
\end{itemize}
\begin{algorithmic}[1] 
\REPEAT
\FOR{$t\mbox{-}$th training iteration}
\STATE $\mathcal{L}_{0}=\mathcal{L}_{rec}+\lambda\mathcal{L}_{cl1}+\beta\mathcal{L}_{cl2}+\gamma\cdot \mathcal{R}$
\STATE $\theta \leftarrow \theta - l\bigtriangleup_{\theta}\mathcal{L}_{0}$
\STATE Update encoder $f_{\theta}(\cdot)$ by minimizing $\mathcal{L}_{0}$
\ENDFOR
\FOR{$t\mbox{-}$th training iteration}
\STATE $\mathcal{L}_{1}= \mathcal{L}_{cl2}+\gamma\cdot \mathcal{R}$
\STATE $\phi1 \leftarrow \phi1 - l'\bigtriangleup_{\phi1}\mathcal{L}_{1}$
\STATE $\phi2 \leftarrow \phi2 - l'\bigtriangleup_{\phi2}\mathcal{L}_{1}$
\STATE Update $w_{\phi1}(\cdot)$ and $w_{\phi2}(\cdot)$ by minimizing $\mathcal{L}_{1}$
\ENDFOR
\UNTIL $\theta,\phi1,\phi2$ converge
\end{algorithmic}
\end{algorithm}
With this meta-learning paradigm, the difference between the dimensions of learned views is more significant, and such informative and discriminative features promote the effectiveness of contrastive learning. In addition, the two modules (i.e., encoder and augmenters) are tightly coupled. More details can be seen in~\ref{rq3} of the experimental section.
\subsection{Contrastive Regularization} 
To prevent creating collapsed augmented views and avoid two augmenters generating too similar contrastive pairs, we further propose a contrastive regularization within updating parameters~\cite{MetAug}. Given two augmented views $\Tilde{\mathbf{z}}^{1}$ and $ \Tilde{\mathbf{z}}^{2}$, we calculate the similarity scores between them by the inner product and then split the output into positive and negative score sets, $\sigma^{+}$ and $\sigma^{-}$, which is calculated as:
\begin{equation}
\sigma^{+},\sigma^{-}=contrast(\Tilde{\mathbf{z}}^{1},\Tilde{\mathbf{z}}^{2}),
\label{eq10}
\end{equation}
where $contrast$ represents the inner product and split operations. The scores calculated from the same sequence are split into positive score sets $\sigma^{+}$, and others are split into negative score sets $\sigma^{-}$. After that, the following formula is used to calculate the regularization:
\begin{equation}
\begin{aligned}
o_{min}&=\min(\{\min(\sigma^{+}),\max(\sigma^{-})\}),\\
o_{max}&=\max(\{\min(\sigma^{+}),\max(\sigma^{-})\}),
\end{aligned}
\label{eq11}
\end{equation}
and
\begin{equation}
\mathcal{R}=\frac{1}{|\sigma^{+}|}\sum([\sigma^{+}-o_{min}]_{+})+\frac{1}{|\sigma^{-}|}\sum([o_{max}-\sigma^{-}]_{+}),
    \label{eq12}
\end{equation}
where $|\sigma^{+}|$ represents the number of positive samples and $|\sigma^{-}|$ represents the number of negative samples. $[\cdot]_{+}$ denotes the cut-off-at-zero function, which is defined as $[a]_{+}=  \max(a, 0)$.
Then the Eq.~(\ref{eq5}) can be rewritten as follows:
\begin{equation}
\mathcal{L}_{0}=\mathcal{L}_{rec}+\lambda\mathcal{L}_{cl1}+\beta\mathcal{L}_{cl2}+\gamma\mathcal{R}.
    \label{eq13}
\end{equation}
The Eq.~(\ref{eq9}) can be rewritten as follows:
\begin{equation}
\mathcal{L}_{1}=\mathcal{L}_{cl2}+\gamma\mathcal{R},
    \label{eq14}
\end{equation}
where $\gamma$ is a weight to balance the contrastive regularization and other losses. The whole training process is detailed by Algorithm~\ref{alg}.

\subsection{Discussion}
\subsubsection{\textbf{Connections with Contrastive SSL in SR}} Recent methods~\cite{CL4SRec,CoSeRec,LMA4Rec,ICLRec,DuoRec,SRMA} mainly take the contrastive objective as an auxiliary task to complement the main recommendation task. 
\begin{table}[t]
  \centering
  \caption{Statistical information of experimented datasets.}
      \renewcommand{\arraystretch}{1}
  \resizebox{1.0\linewidth}{!}{
    \begin{tabular}{cccccc}
    \hline
    Datasets & \#users & \#items & \#actions & avg.length & sparsity \\
    \hline
    Sports & 35598 & 18357 & 296337 & 8.3   & 99.95\% \\
    Beauty & 22363 & 12101 & 198502 & 8.8   & 99.93\% \\
    Yelp  & 30431 & 20033 & 316354 & 10.4  & 99.95\% \\
    \hline
    \end{tabular}}%
  \label{tab:addlabel}%
\end{table}%
Among them, CL4SRec~\cite{CL4SRec}, CoSeRec~\cite{CoSeRec}, and ICLRec~\cite{ICLRec} augment the input sequence at data level with cropping, masking, and reordering. 
DuoRec~\cite{DuoRec} conducts neural masking augmentation on the input sequence at the model level.
LMA4Rec~\cite{LMA4Rec} introduces Learnable Bernoulli Dropout (LBD~\cite{bernoulli}) to the encoder and combines it with stochastic data augmentation to construct contrastive views. 
SRMA~\cite{SRMA} introduces three model augmentation methods (neural masking, layer dropping, and encoder complementing) and combines them with data augmentation for constructing view pairs. 
However, all the above models use either stochastic data augmentation or stochastic model augmentation. 
Different from the above-mentioned models, our model can be viewed as a two-stage process combining both data and model augmentation operations. In the first stage, stochastic data augmentation is applied to obtain two pairwise contrastive views, and adaptively learn more informative features from these views by using two learnable augmenters in the second stage. Compared with other CL models, MCLRec leverages data and model augmentation views to enlarge the number of contrastive pairs without increasing the input data, thus extracting more informative features for the model training.
Meanwhile, the meta-learning optimized approach is also implemented to guide the training of learnable augmenters, which is an alternative way to fuse these two augmentation manners. The main differences are summarized in Table~\ref{tab:contrast}.  
\subsubsection{\textbf{Time Complexity Analysis of MCLRec}} 
The complexity of our model mainly comes from the training and the testing. During training, the computation costs of our proposed method are mainly from the optimization of $\theta$, $\phi_{1}$, and $\phi_{2}$ with multi-task learning in two stages. For stage one, since we have four objectives to optimize the network $f_{\theta}$, the time complexity is $\mathcal{O}(|\mathcal{U}|^{2}d+|\mathcal{U}|d^{2})$. 

For stage two, we have two objectives to optimize the augmenters, the time complexity is $\mathcal{O}(|\mathcal{U}|d^{2})$. The overall complexity is dominated by the term $\mathcal{O}(|\mathcal{U}|^{2}d)$, where $|\mathcal{U}|$ represents the number of users. 
In the testing phase, the proposed augmenters and contrastive objectives are no longer needed, which enables the model to have the time complexity as the encoder, e.g., SASRec~$(\mathcal{O}(d|\mathcal{I}|))$, where $|\mathcal{I}|$ represents the number of items.  
Based on the above analysis, our MCLRec achieves comparable time complexity when computing with state-of-the-art contrastive SR models~\cite{CL4SRec,ICLRec}.
\section{Experiment}
In this section, we conduct extensive experiments with three real-world datasets, investigating the following research questions (\textbf{RQ}s). 
\begin{itemize}[leftmargin=*]
    \item \textbf{RQ1}: How does MCLRec perform compared to state-of-the-art sequential recommendation models? 
    \item \textbf{RQ2}: How effective are the key model components (e.g., stochastic data augmentation, learnable model augmentation, contrastive regularization) in MCLRec?
    \item \textbf{RQ3}: How does the meta-learning training strategy affect the recommendation performance?
    \item \textbf{RQ4}: How does the robustness (e.g., training on small batch size, adding noise on test datasets) of MCLRec? 
\end{itemize}
\begin{table*}[t]
\centering
\caption{Performance comparisons of different methods. Where the bold score is the best in each row and the second-best baseline is underlined. The last column is the relative improvements compared with the best baseline results.}
\renewcommand{\arraystretch}{1.0}
    \resizebox{\textwidth}{!}{
  \begin{tabular}{c|l|c|ccc|cccccccc||cc}
    \hline
    Dataset&
     Metric &  BPR &   GRU4Rec &  Caser&   SASRec&
     BERT4Rec & S$^3$-Rec$_{\rm{MIP}}$& CL4SRec& CoSeRec&LMA4Rec&
     ICLRec&DuoRec &SRMA&
    MCLRec& Improv. \\
   \hline
  \hline
    \multirow{6}*{Sports}& HR@5 & 0.0123 & 0.0162 & 0.0154 & 0.0214 & 0.0217 & 0.0121 & 0.0231 &0.0290&0.0297 &0.0290 & \underline{0.0312}& 0.0299&\textbf{0.0328}& 5.13\%\\
    
    &HR@10&0.0215&0.0258&0.0261&0.0333&0.0359&0.0205&0.0369&0.0439&0.0439&0.0437&\underline{0.0466}&0.0447&\textbf{0.0501}&7.51\%\\
    
    & HR@20& 0.0369 & 0.0421 & 0.0399 & 0.0500 & 0.0604 & 0.0344 & 0.0557  &0.0636&0.0634&0.0646 &\underline{0.0696}& 0.0649&\textbf{0.0734}& 5.46\%\\
    
    & NDCG@5 & 0.0076 & 0.0103 & 0.0114 & 0.0144 & 0.0143 & 0.0084 & 0.0146 &0.0196&0.0197&0.0191&0.0192 &\underline{0.0199}& \textbf{0.0204}& 2.51\%\\
    
    &NDCG@10 &0.0105&0.0142&0.0135&0.0177&0.0190&0.0111&0.0191&0.0244&0.0245&0.0238&0.0244&\underline{0.0246}&\textbf{0.0260} &5.69\%\\
    
    & NDCG@20 & 0.0144 & 0.0186 & 0.0178 & 0.0224 & 0.0251 & 0.0146 & 0.0238  &0.0293&0.0293&0.0291&\underline{0.0302}&0.0297 &\textbf{0.0319}& 5.63\%\\
    \hline
    
    \multirow{6}*{Beauty}& HR@5 & 0.0178 & 0.0180 & 0.0251 & 0.0377 & 0.0360 & 0.0189 & 0.0401  &0.0504&0.0511&0.0500& \underline{0.0559}&0.0503  &\textbf{0.0581} & 3.94\%\\
    
    &HR@10& 0.0296&0.0284&0.0342&0.0624&0.0601&0.0307&0.0642&0.0725&0.0735&0.0744&\underline{0.0825}&0.0724&\textbf{0.0871} & 5.58\%\\
    
    & HR@20& 0.0474 & 0.0427 & 0.0643 & 0.0894 & 0.0984 & 0.0487 & 0.0974  &0.1034&0.1047&0.1058&\underline{0.1193}&0.1025  &\textbf{0.1243} &4.19\%\\
    
    & NDCG@5 & 0.0109 & 0.0116 & 0.0145 & 0.0241 & 0.0216& 0.0115 & 0.0268  &0.0339&\underline{0.0342}&0.0326&0.0340&0.0318&\textbf{0.0352}& 2.92\%\\
    
    &NDCG@10 &0.0147&0.0150&0.0226&0.0342&0.0300&0.0153&0.0345&0.0410&0.0414&0.0403&\underline{0.0425} &0.0398&\textbf{0.0446}& 4.94\%\\
    
    & NDCG@20 & 0.0192 & 0.0186 & 0.0298 & 0.0386 & 0.0391 & 0.0198 & 0.0428  &0.0487 &0.0493&0.0483&\underline{0.0518} &0.0474 &\textbf{0.0539}& 4.05\%\\
    \hline
    
    \multirow{6}*{Yelp}& HR@5 & 0.0127 & 0.0152 & 0.0142 & 0.0160 & 0.0196 & 0.0101 & 0.0227  &0.0241&0.0233&0.0239& \underline{0.0429}  &0.0243& \textbf{0.0454} & 5.83\%\\
    
    &HR@10&0.0216&0.0248&0.0254& 0.0260&0.0339&0.0176&0.0384&0.0395&0.0387&0.0409&\underline{0.0614}&0.0395&\textbf{0.0647}& 5.37\% \\
    
    & HR@20& 0.0346 & 0.0371 & 0.0406 & 0.0443 & 0.0564 & 0.0314 & 0.0623 &0.0649&0.0636 &0.0659&\underline{0.0868} &0.0646 & \textbf{0.0941} & 8.41\%\\
    
    & NDCG@5 & 0.0082 & 0.0091 & 0.0080 & 0.0101 & 0.0121 & 0.0068 & 0.0143  &0.0151&0.0147&0.0152&\underline{0.0324}  &0.0154& \textbf{0.0332}& 2.47\%\\
    
    &NDCG@10&0.0111&0.0124&0.0113&0.0133&0.0167&0.0092&0.0194&0.0205&0.0196&0.0207&\underline{0.0383}&0.0207&\textbf{0.0394}& 2.87\%\\
    
    & NDCG@20 & 0.0143 & 0.0145 & 0.0156 & 0.0179 & 0.0223 & 0.0127 & 0.0254  &0.0263&0.0258&0.0270& \underline{0.0447}&0.0266 & \textbf{0.0467} & 4.47\%\\
    \hline
\end{tabular}}
  \label{tabr}
\end{table*}
\subsection{Experimental Settings}
\subsubsection{\textbf{ Datasets}.} To verify the effectiveness of our methods, we evaluate the model on three real-world benchmark datasets: Amazon (Beauty and Sports)\footnote{\url{https://jmcauley.ucsd.edu/data/amazon/}} and Yelp\footnote{\url{https://www.yelp.com/dataset}}. Amazon dataset collects user review data from amazon.com, which is one of the largest e-commerce websites in the world. We use two sub-categories, Amazon-Beauty and Amazon-Sports, in our experiments. Yelp is a dataset for business recommendations. Following~\cite{S3Rec,CoSeRec,DuoRec} for preprocessing, the users and items that have less than five interactions are removed.
The statistics of the prepared datasets are summarized in Table~\ref{tab:addlabel}.  
\subsubsection{\textbf{ Baseline Methods}.} 
We compare our models with the following three groups of sequential recommendation models:
\begin{itemize}[leftmargin=*]
\item\textbf{Non-sequential models}: 
BPR~\cite{BPR} uses Bayesian Personalized Ranking (BPR) loss to optimize the matrix factorization model. 

\item \textbf{General sequential models}: GRU4Rec~\cite{GRU4Rec} uses Gated Recurrent Unit (GRU) to model for the sequential recommendation. Caser~\cite{Caser} uses both horizontal and vertical Convolution Neural Networks (CNN) to model sequential behaviors. 
SASRec~\cite{SASRec} for the first time to use the attention mechanism to the sequential recommendation and achieve a good performance.

\item \textbf{Self-supervised based sequential models}: BERT4Rec~\cite{BERT4Rec} uses the deep bidirectional self-attention to capture the potential relationships between items and sequences in Cloze task~\cite{Cloze}. 
S$^{3}$-Rec~\cite{S3Rec} uses self$\mbox{-}$supervised learning to capture the correlations between items. Since there is no attribute information in our experiments, only the MIP (Masked Item Prediction) task, called S$^{3}$-Rec$_{\rm{MIP}}$, is used for training.
CL4SRec~\cite{CL4SRec} uses both data augmentation and contrastive learning in the sequential recommendation for the first time. 
CoSeRec~\cite{CoSeRec} further proposes two more informative data augment methods (i.e., `insert' and `substitute') to improve the performance of contrastive learning. 
LMA4Rec~\cite{LMA4Rec} improves CoSeRec by introducing a Learnable Bernoulli Dropout (LBD ~\cite{bernoulli}) to the encoder, which is to  extract more signals from the stochastic augmented views. 
ICLRec~\cite{ICLRec} learns users' latent intents from the behavior sequences through clustering and integrates the learned intents into the model via an auxiliary contrastive SSL loss. 
DuoRec~\cite{DuoRec} proposes a sampling strategy to formulate positive samples and uses dropout~\cite{dropout} to conduct the model-level augmentation. 
SRMA~\cite{SRMA} introduces three model augmentation methods (i.e., `neural mask', `layer drop', and `encoder complement') and combines them with data augmentation for constructing view pairs.
\end{itemize}

\subsubsection{\textbf{Evaluation Metrics}.}  For evaluation purposes, we split the data into training, validation, and testing datasets based on timestamps given in the datasets~\cite{SASRec, ICLRec, DuoRec}. Specifically, the last item is used for testing, the second-to-last item is used for validation, and the rest for training.
Following ~\cite{Metric2,Metric1}, we rank the whole item set without negative sampling. In order to evaluate the model effectively, we use two widely$\mbox{-}$used evaluation metrics, including Hit Ratio @$k$ (HR@$k$) and Normalized Discounted Cumulative Gain @$k$ (NDCG@$k$), where $k \in \{5,10,20\}$. Intuitively, the HR metric considers whether the ground$\mbox{-}$truth is ranked amongst the top $k$ items while the NDCG metric is a position$\mbox{-}$aware ranking metric.

\subsubsection{\textbf{Implementation Details}.}
The implementations of Caser, S$^{3}$-Rec, BERT4Rec, CoSeRec, LMA4Rec, ICLRec, DuoRec, and SRMA are provided by the authors. BPR, GRU4Rec, SASRec, and CL4SRec are implemented based on public resources. All parameters in these methods are used as reported in their papers and the optimal settings are chosen based on the model performance on validation data. For MCLRec, we use transformer~\cite{SASRec} as the encoder, and the number of the self-attention blocks and attention heads is set as 2. The augmenters are 3-layer fully connected MLPs. We set $d$ as 64, $n$ as 50, the learning rate $l$ as 0.001, $l'$ as 0.001, and the batch size as 256. $\lambda$, $\beta$ are selected from $\{0.01, 0.02, 0.03, 0.04, 0.05, 0.1, 0.2, 0.3, 0.4, 0.5\}$ and $\gamma$ is $0.1 \times \beta$. The whole model is optimized with the Adam~\cite{adam} optimizer. 
We train the model with an early stopping strategy based on the performance of validation data. All experiments are implied on NVIDIA GeForce RTX 2080 Ti GPU.

\subsection{Overall Performances (RQ1)}
We compare the performance of all baselines with MCLRec for Sequential Recommendation. Table~\ref{tabr} shows the experimental results of the compared models on three datasets, and the following findings can be seen through it:
\begin{itemize}[leftmargin=*]
\item The self-supervised based models perform more effectively than classical models, such as BPR, GRU4Rec, Caser, and SASRec. Among them, different from BERT4Rec and S$^{3}$-Rec$_{\rm{MIP}}$ that use MIP tasks to train the model, CL4SRec, CoSeRec, LMA4Rec, ICLRec, DuoRec, and SRMA utilize data augmentation and contrastive learning for training, which lead to generally better results than BERT4Rec and S$^{3}$-Rec$_{\rm{MIP}}$. That indicates contrastive learning paradigm may generate more expressive embeddings for users and items by maximizing the mutual information.
\item Compared to SRMA and CL4SRec, we can find that introducing model augmentation can further improve performance. In addition, DuoRec performs better than other baselines on all datasets. Compared with the previous SSL-based sequential models, DuoRec utilizes both supervised data augmentation and random model augmentation, and thus improves the performance by a large margin. That motivates us to combine two types of augmentation operations within the training of CL. 
\item Benefiting from the meta-optimized model augmentation operation, MCLRec significantly outperforms other methods on all metrics across the different datasets. For instance, MCLRec improves over the second-best result w.r.t. HR and NDCG by 3.94-8.41\% and 2.47-5.69\% on three datasets, respectively. The reasons are concluded as: (1) Our learnable augmentation module adaptively learns appropriate augmentation representations for contrastive learning. (2) The meta-learning manner acts as an effective way for training the augmentation model as well as boosting recommendation accuracy. The results support that our contrastive recommendation framework can enable different models to learn more informative representations. 
\end{itemize}
\begin{figure}[t]
\centering
    \begin{minipage}[h]{0.49\linewidth}
        \centering
        \centerline{\includegraphics[width=1.0\textwidth]{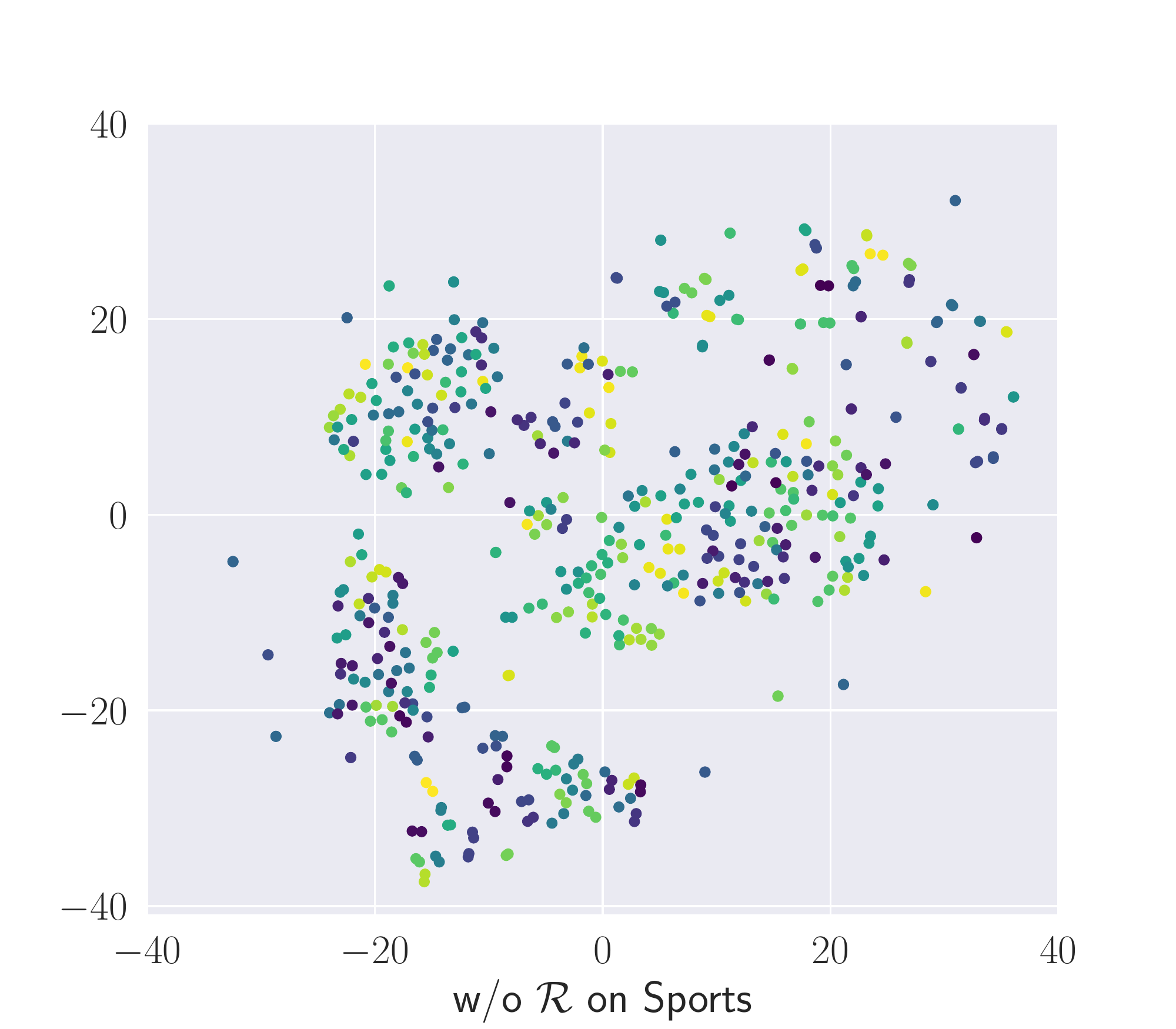}}
    \end{minipage}
    \begin{minipage}[h]{0.49\linewidth}
        \centering
        \centerline{\includegraphics[width=1.0\textwidth]{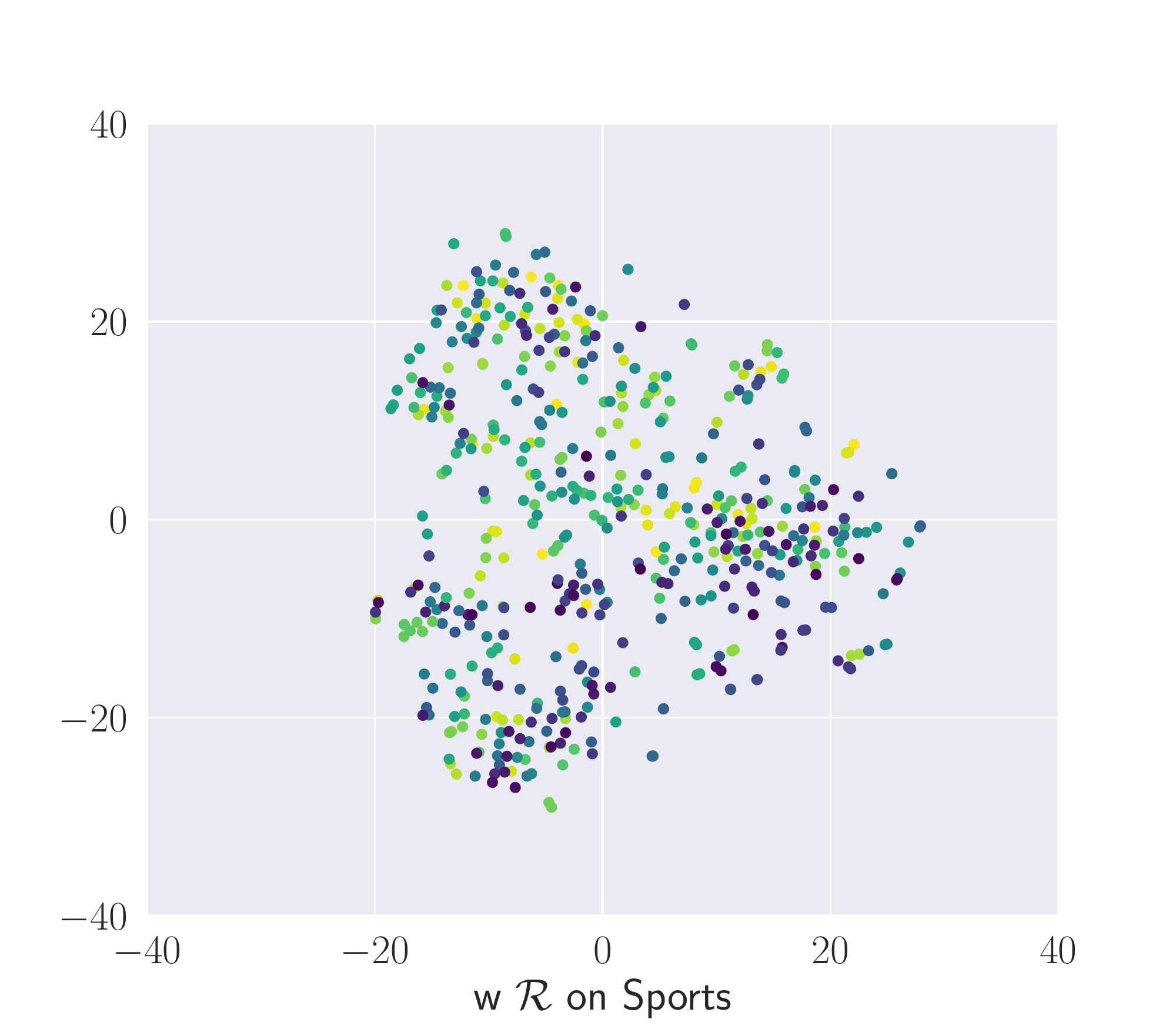}}
    \end{minipage}
    \begin{minipage}[h]{0.49\linewidth}
        \centering
        \centerline{\includegraphics[width=1.0\textwidth]{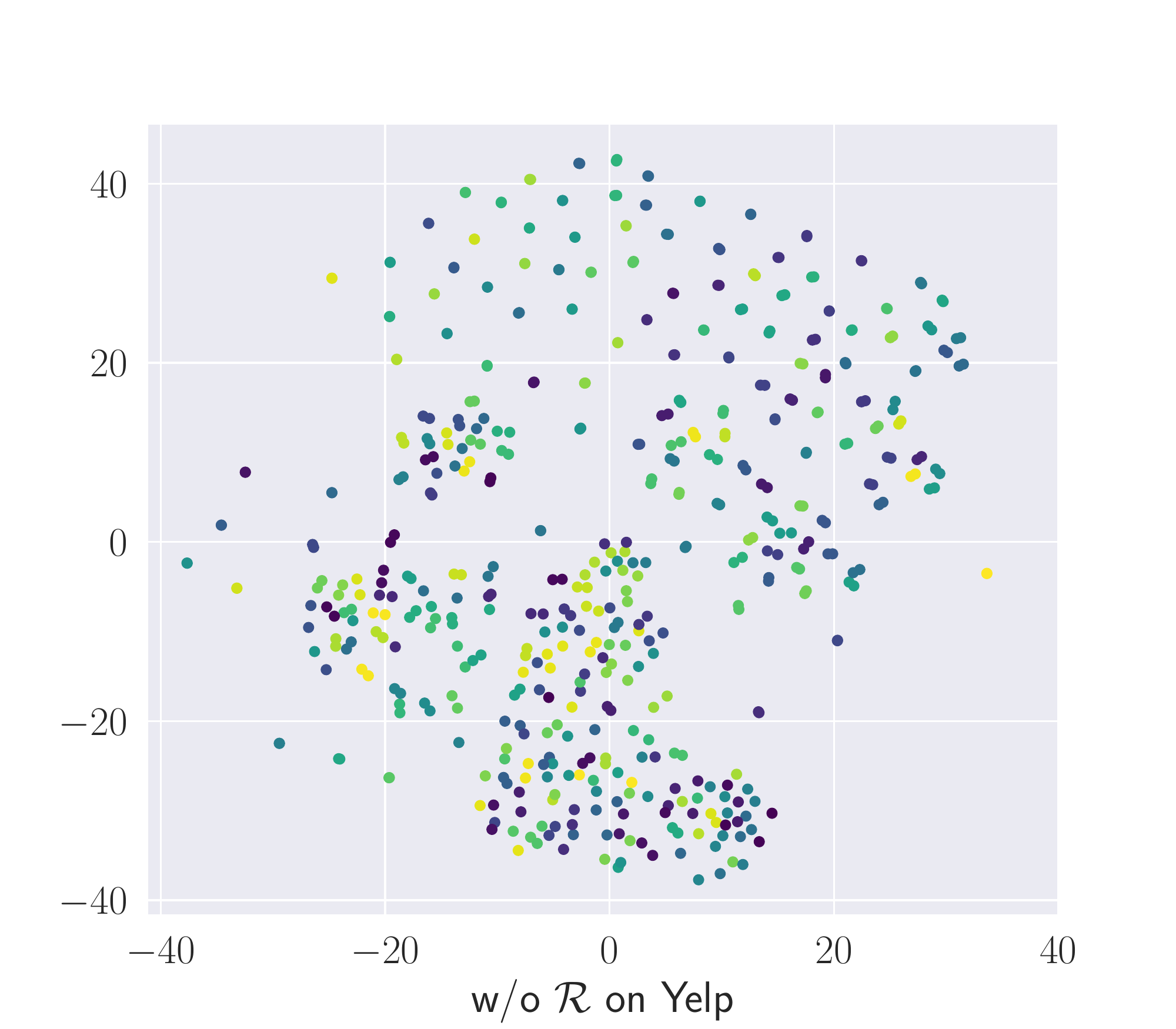}}
    \end{minipage}
    \begin{minipage}[h]{0.49\linewidth}
        \centering
        \centerline{\includegraphics[width=1.0\textwidth]{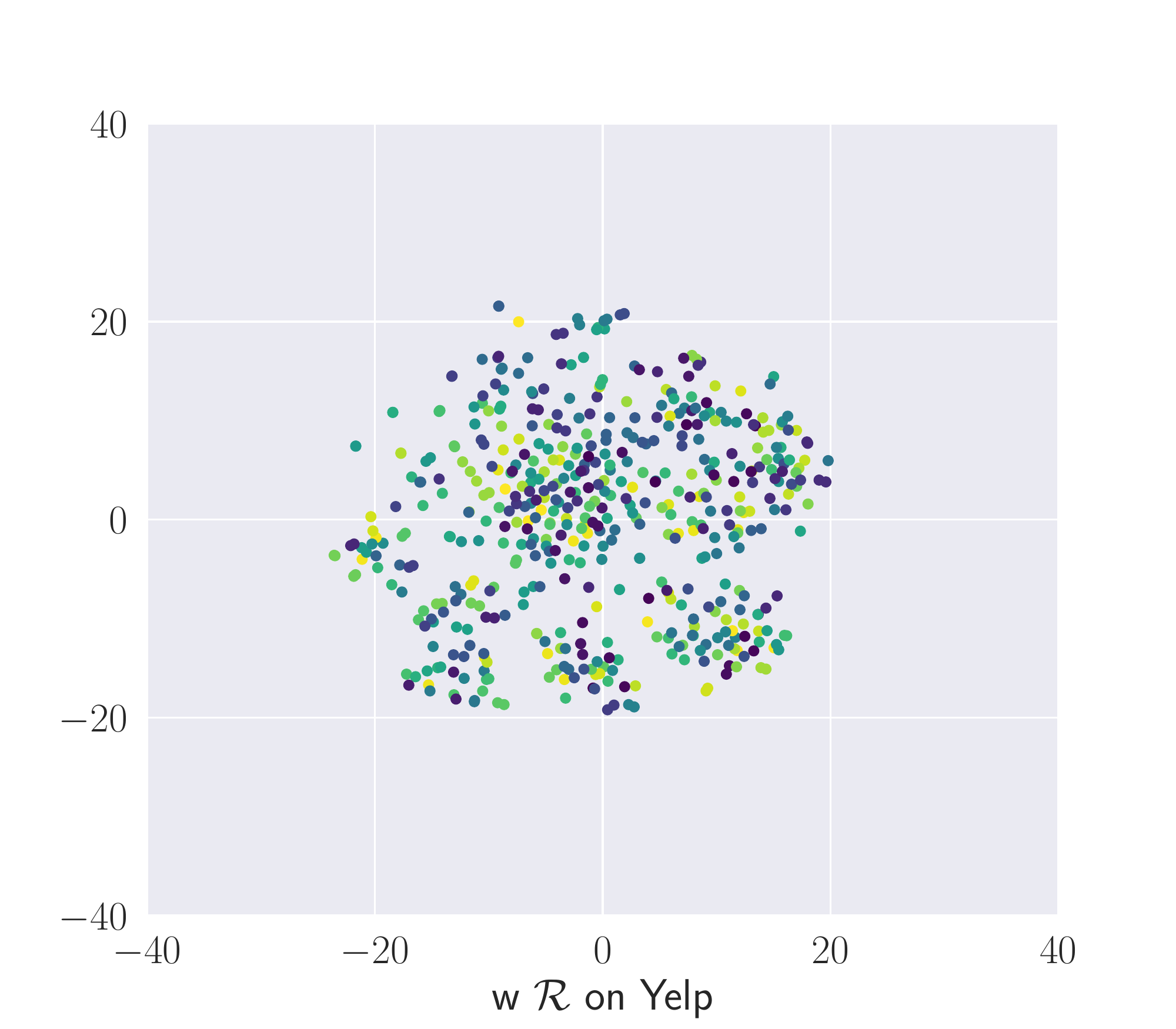}}
    \end{minipage}
       \caption{T-SNE visualization of the model augmentation views $\Tilde{\mathbf{z}}^{1}$ and $\Tilde{\mathbf{z}}^{2}$ trained with w/o $\mathcal{R}$ and w $\mathcal{R}$ on Sports and Yelp. Where different colors represent negative pairs.}
         \label{rl_exp}
\end{figure} 

\begin{table}[t]
  \centering
      \caption{Ablation study with key modules. Where HR and NDCG indicate HR@20 and NDCG@20.}
  \renewcommand{\arraystretch}{1}
  \resizebox{1.0\linewidth}{!}{
    \begin{tabular}{lrr|cccc|cccc|cccc}
    \hline
    \multicolumn{3}{c|}{\multirow{3}[4]{*}{Model}} & \multicolumn{12}{c}{Dataset} \\
\cline{4-15}\multicolumn{3}{c|}{} & \multicolumn{4}{c|}{Sports}& \multicolumn{4}{c|}{Beauty}&\multicolumn{4}{c}{Yelp} \\
    \multicolumn{3}{c|}{} & \multicolumn{2}{c}{HR} & \multicolumn{2}{c|}{NDCG}&\multicolumn{2}{c}{HR} & \multicolumn{2}{c|}{NDCG} & \multicolumn{2}{c}{HR} & \multicolumn{2}{c}{NDCG} \\
    \hline
    \multicolumn{3}{l|}{          (A) MCLRec} & \multicolumn{2}{c}{\textbf{0.0734}} & \multicolumn{2}{c|}{\textbf{0.0319}}&\multicolumn{2}{c}{\textbf{0.1243}} & \multicolumn{2}{c|}{\textbf{0.0539}} & \multicolumn{2}{c}{\textbf{0.0941}} & \multicolumn{2}{c}{\textbf{0.0467}} \\
    \multicolumn{3}{l|}{          (B) w/o $\mathcal{L}_{cl1}$}& \multicolumn{2}{c}{0.0705} & \multicolumn{2}{c|}{0.0299}&\multicolumn{2}{c}{\textbf{0.1243}} & \multicolumn{2}{c|}{\textbf{0.0539}} & \multicolumn{2}{c}{0.0918} & \multicolumn{2}{c}{0.0462} \\
    \multicolumn{3}{l|}{          (C) w/o $\mathcal{L}_{cl2}$} & \multicolumn{2}{c}{0.0557} & \multicolumn{2}{c|}{0.0238}&\multicolumn{2}{c}{0.1056} & \multicolumn{2}{c|}{0.0394} & \multicolumn{2}{c}{0.0623} & \multicolumn{2}{c}{0.0254} \\
    \multicolumn{3}{l|}{          (D) w/o $\mathcal{R}$} & \multicolumn{2}{c}{0.0691} & \multicolumn{2}{c|}{0.0291}&\multicolumn{2}{c}{0.1236} & \multicolumn{2}{c|}{0.0529} & \multicolumn{2}{c}{0.0873} & \multicolumn{2}{c}{0.0445} \\
      \multicolumn{3}{l|}{          (E) share}& \multicolumn{2}{c}{0.0707} & \multicolumn{2}{c|}{0.0299}&\multicolumn{2}{c}{0.1231} & \multicolumn{2}{c|}{0.0532} & \multicolumn{2}{c}{0.0923} & \multicolumn{2}{c}{0.0456} \\
      \hline
    \end{tabular}%
    }
  \label{tab:3}%
\end{table}%

\subsection{Ablation Study (RQ2)}
To analyze the effectiveness of each component of our model, we conduct several ablation experiments about MCLRec. HR@20 and NDCG@20 performances of different variants are shown in Table~\ref{tab:3}, where w/o denotes without, (A) represents MCLRec, (B) removes the $\mathcal{L}_{cl1}$ by setting $\lambda$ to 0 in Eq.~(\ref{eq13}), (C) removes the $\mathcal{L}_{cl2}$ (, which is equivalent to CL4SRec), (D) removes the contrastive regularization component, and (E) denotes the two augmenters that share parameters (i.e., $\phi1 = \phi2$). From this table, we can find that MCLRec achieves the best results on all datasets, which indicates all components are effective for our framework and the meta-optimized contrastive learning enhances the model's ability to learn more expressive representations. 
By comparing (A) with (C) and (D), we find that learnable model augmentation and contrastive regularization could significantly improve the model accuracy, which is consistent with our statements.
By comparing (B) and (C), it can be observed that learnable augmentation is much more efficient than random data augmentation.
By comparing (A) and (B), the combination of data and model level augmentation could further boost model performance.
By comparing (A) and (E), we can find that sharing parameters of augmenters will decrease the results. This may be the fact that using the same augmenter may further lead to a high similarity of learned augmentation views, thus making the performance degraded.
As shown in Table~\ref{tab:3}, after removing the regular term, the performance of our model decreases on all three datasets, which indicates the effectiveness of the regular term. 
\begin{figure}[t]
\centering
    
      \begin{minipage}[t]{0.49\linewidth}
        \centering
        \includegraphics[width=1.0\textwidth]{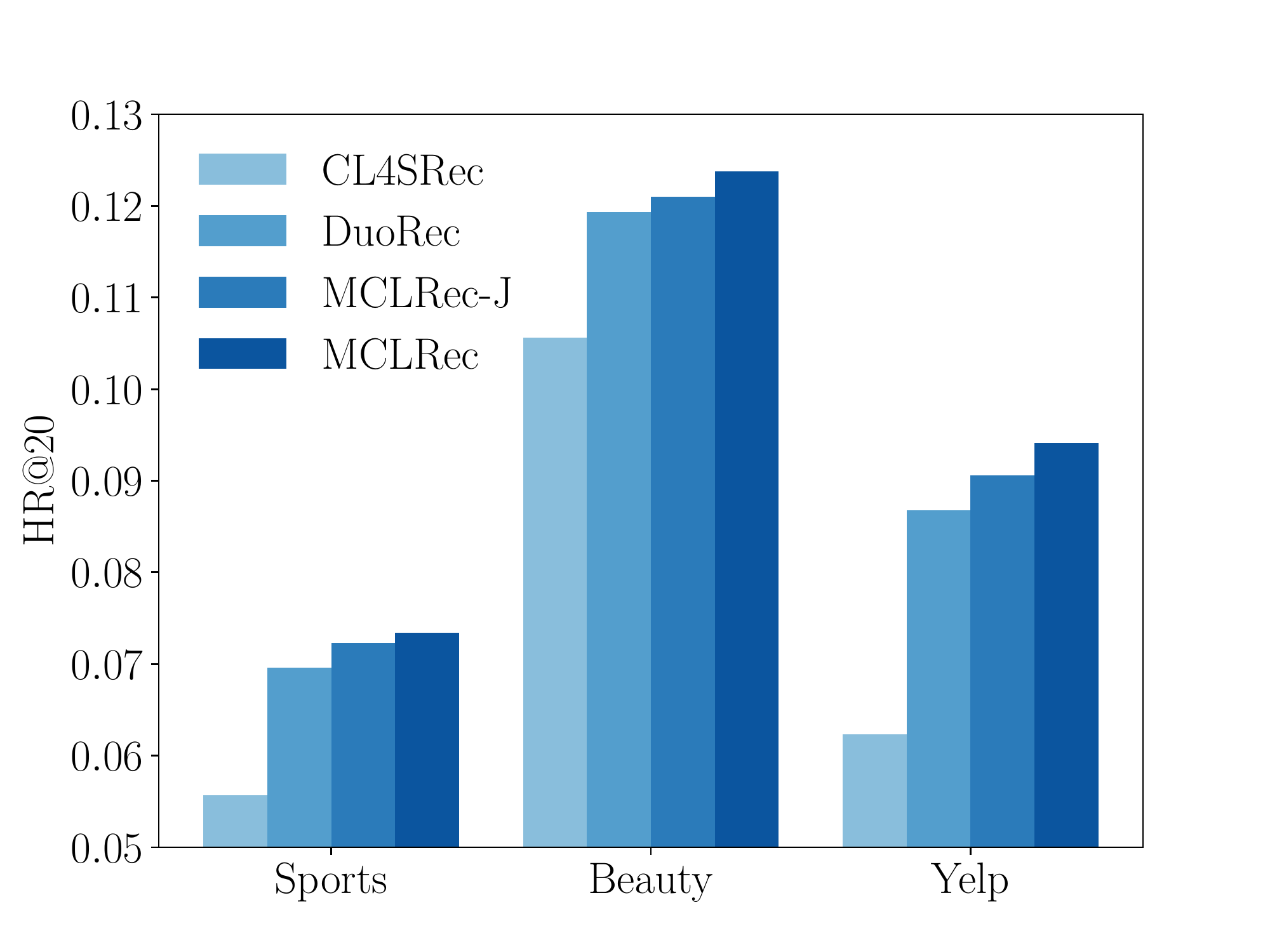}
    \end{minipage}
        \begin{minipage}[t]{0.49\linewidth}
        \centering        \includegraphics[width=1.0\textwidth]{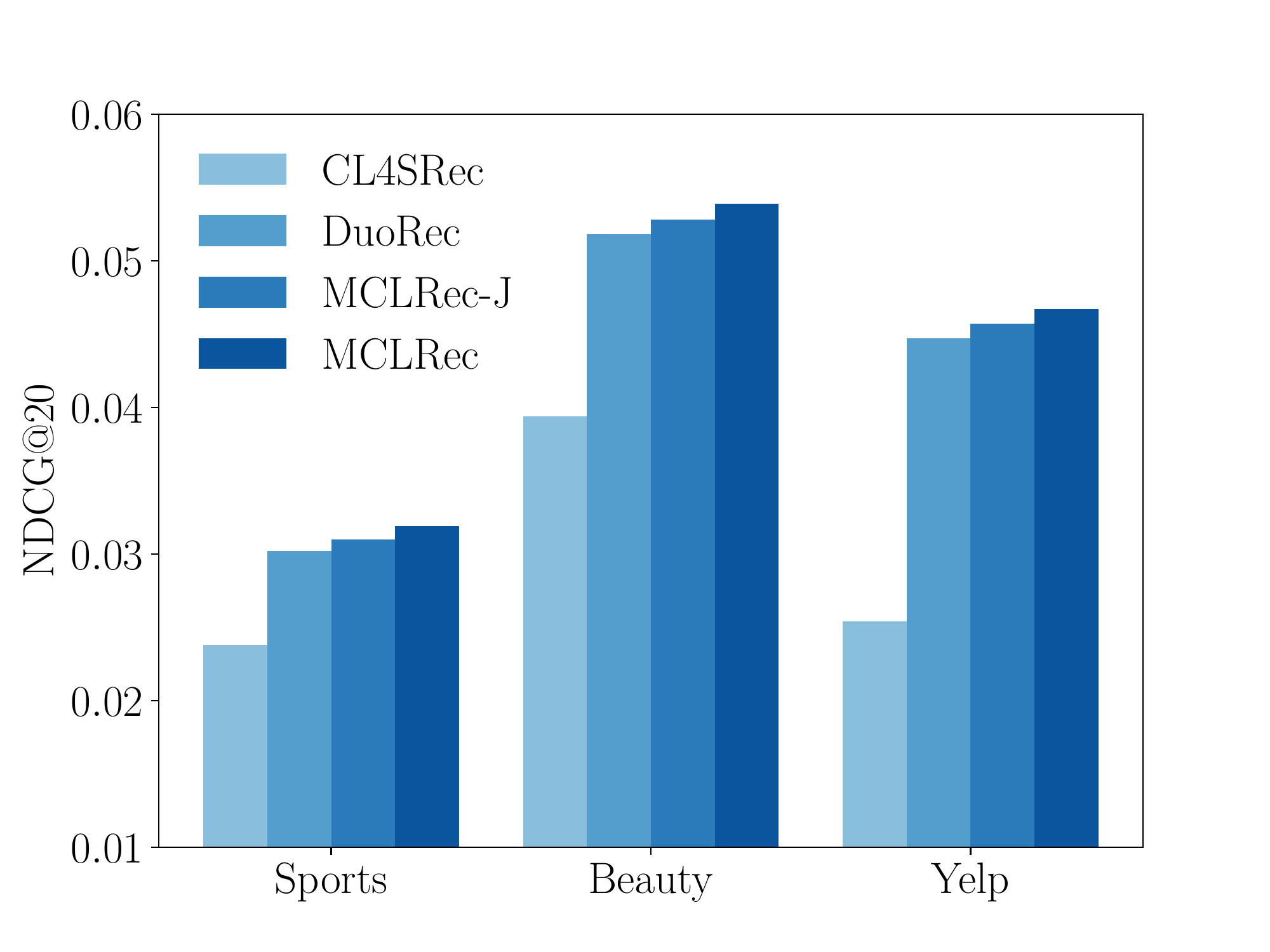}
    \end{minipage}%
    \caption{Comparison of two versions MCLRec (have different train strategies) with CL4SRec and DuoRec on all datasets.}
    \label{strategy_exp}
    \begin{minipage}[h]{0.49\linewidth}
        \centering
    \centerline{\includegraphics[width=1.0\textwidth]{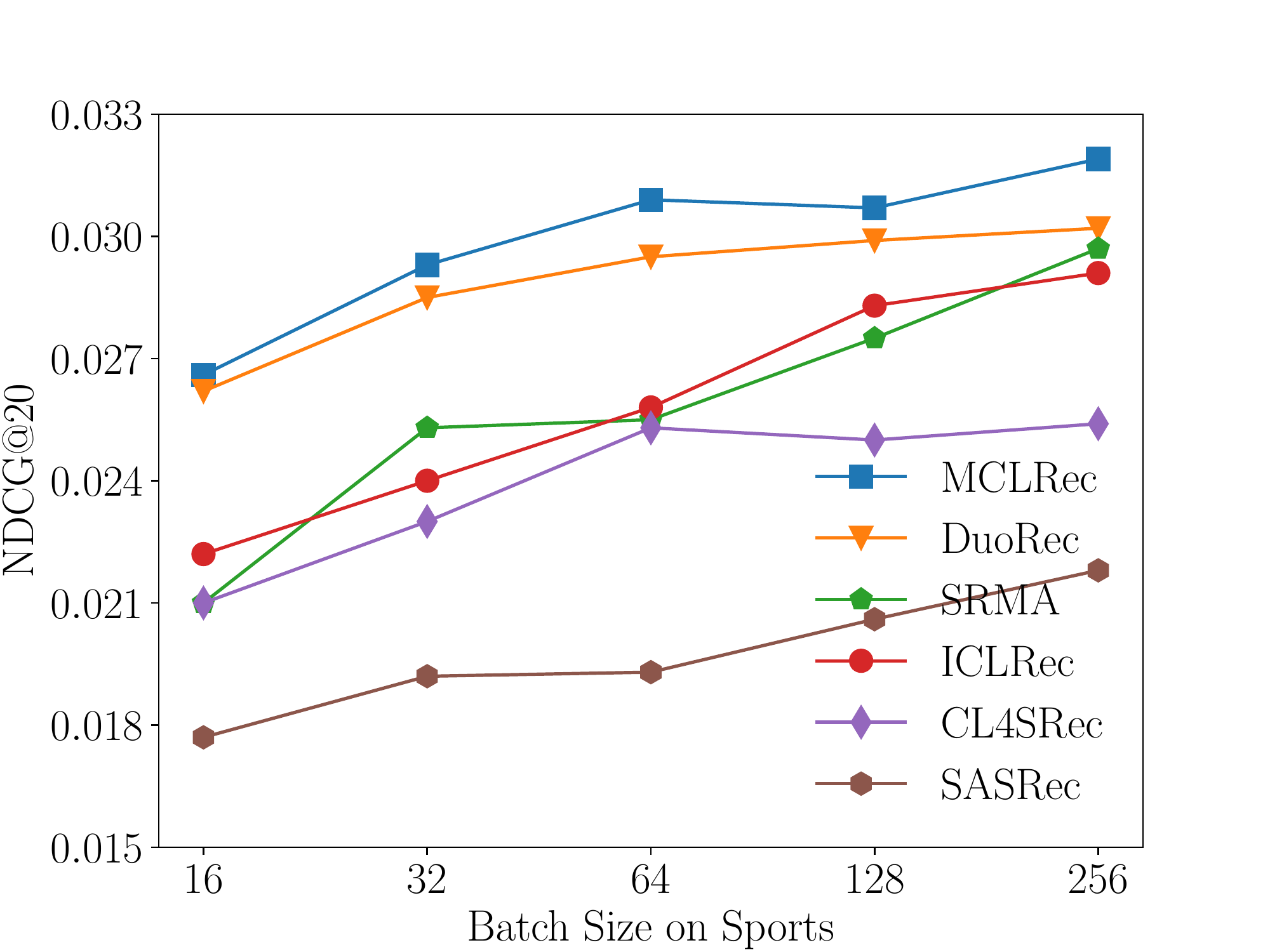}}
    \end{minipage}
    \begin{minipage}[h]{0.49\linewidth}
        \centering
    \centerline{\includegraphics[width=1.0\textwidth]{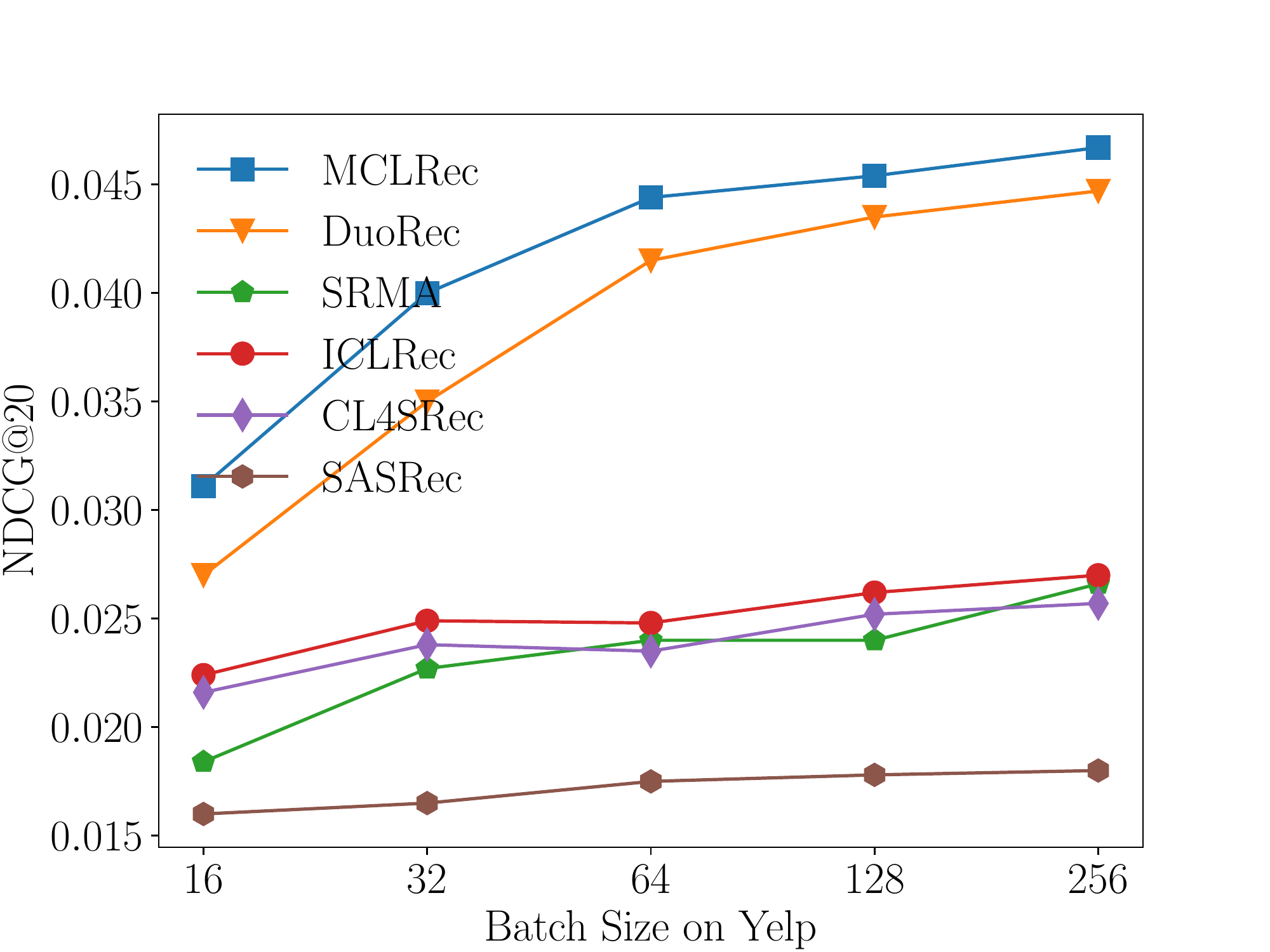}}
    \end{minipage}
      \caption{Performances comparison w.r.t. Batch Size.} 
         \label{batch_exp}
\end{figure}
To further analyze the effect of the regular term on the model, we visualize the learned augmentation views $\Tilde{\mathbf{z}}^{1}$ and $\Tilde{\mathbf{z}}^{2}$ via T-SNE~\cite{t-SNE}. To simplify, we denote without as w/o and with as w. We use both w/o $\mathcal{R}$ and w $\mathcal{R}$ to train our model for 300 epochs in an end-to-end manner respectively and utilize T-SNE to reduce the augmented embeddings into two-dimensional space. Limited by the space, the results of Sports and Yelp are presented in Figure~\ref{rl_exp}. 
We find that w/o $\mathcal{R}$ allows the enhancer to learn collapsed view representations (i.e., the representations of both positive and negative pairs are too "dispersed"), and w $\mathcal{R}$ allows the augmenters to learn more discriminative features (i.e., the positive pairs are "close" enough and the negative pairs are relative "far away"). This further demonstrates the effectiveness of the regular term. In addition, we found that simply adding $\mathcal{R}$ to other models (e.g., CL4SRec and DuoRec) will make the performance worse. The reason might be that $\mathcal{R}$ is mainly designed to constrain model augmenters.  
\subsection{Effectiveness of Meta Optimization (RQ3)}\label{rq3}
We conduct several experiments based on MCLRec to analyze the effectiveness of meta optimization. We first compare the performance of the joint-learning strategy with MCLRec, called MCLRec-J, where the joint-learning strategy means the whole model is trained according to $\mathcal{L}_{0}$ (Eq.~(\ref{eq13})) in one step. As shown in Figure~\ref{strategy_exp}, our meta-optimized based manner outperforms others on all datasets. Specifically, \emph{on the one hand}, MCLRec-J performs better than CL4SRec and DuoRec, which demonstrates the effectiveness of learnable model augmentation. \emph{On the other hand}, MCLRec beats the MCLRec-J. The main reason is that meta-learning help CL to learn more discriminative augmentation views. To specify this point, we visualize the learned augmentation views $\Tilde{\mathbf{h}}^{1}$ and $\Tilde{\mathbf{h}}^{2}$ via T-SNE~\cite{t-SNE}. 
\begin{figure}[t]
\centering
    \begin{minipage}[h]{0.49\linewidth}
        \centering
    \centerline{\includegraphics[width=1.0\textwidth]{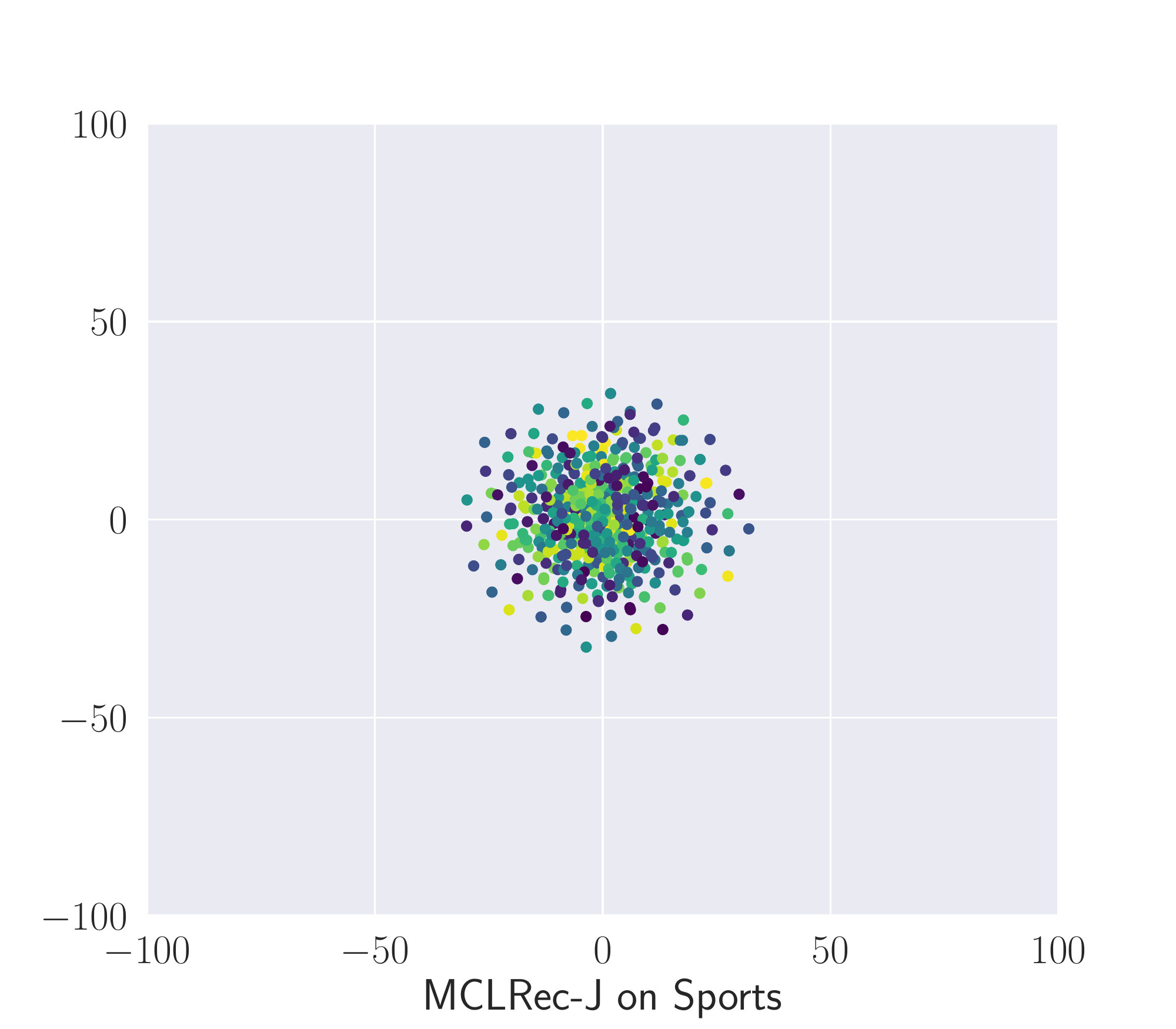}}
    \end{minipage}
    \begin{minipage}[h]{0.49\linewidth}
        \centering
        \centerline{\includegraphics[width=1.0\textwidth]{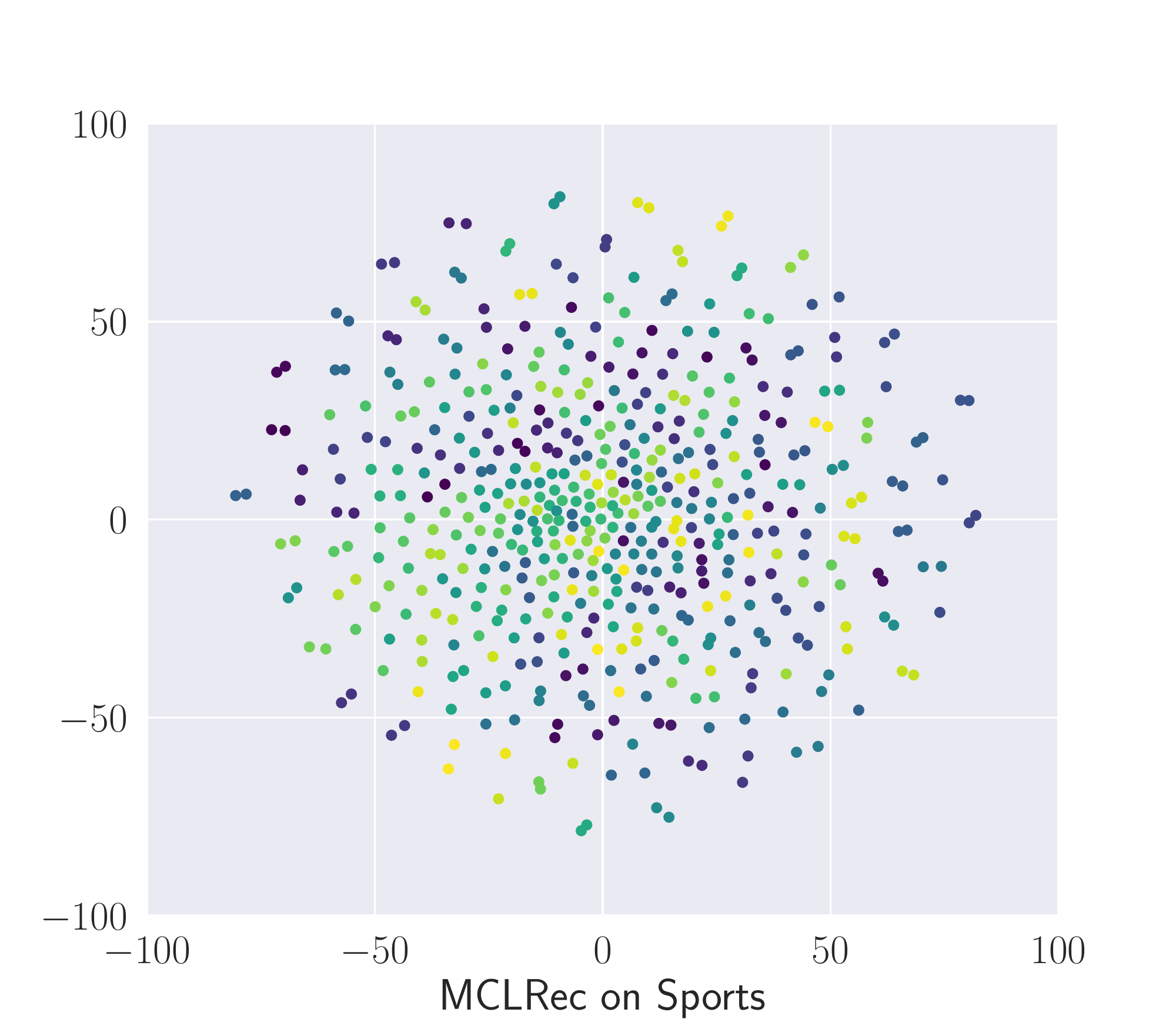}}
    \end{minipage}
    \begin{minipage}[h]{0.49\linewidth}
        \centering
        \centerline{\includegraphics[width=1.0\textwidth]{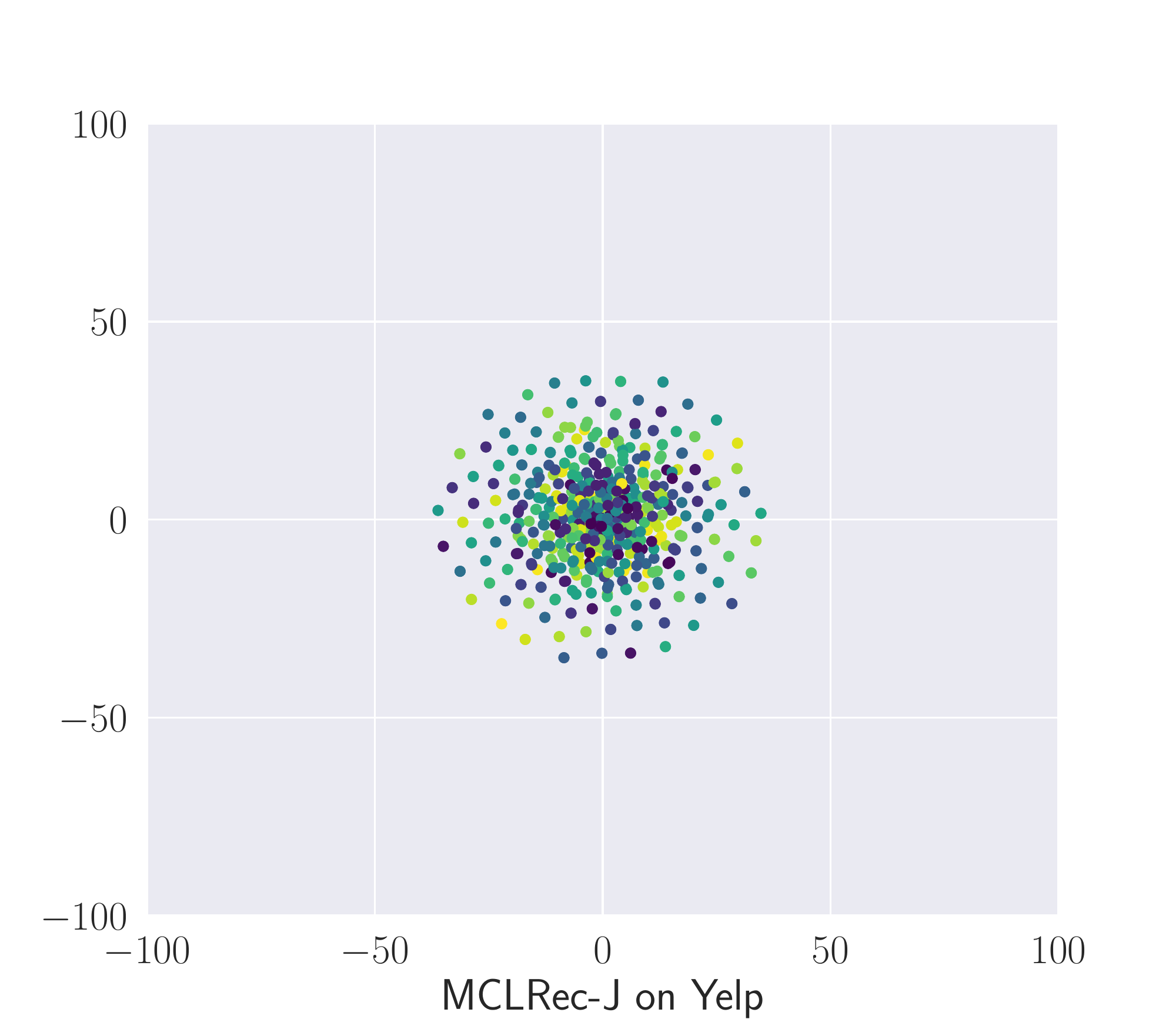}}
    \end{minipage}
    \begin{minipage}[h]{0.49\linewidth}
        \centering
        \centerline{\includegraphics[width=1.0\textwidth]{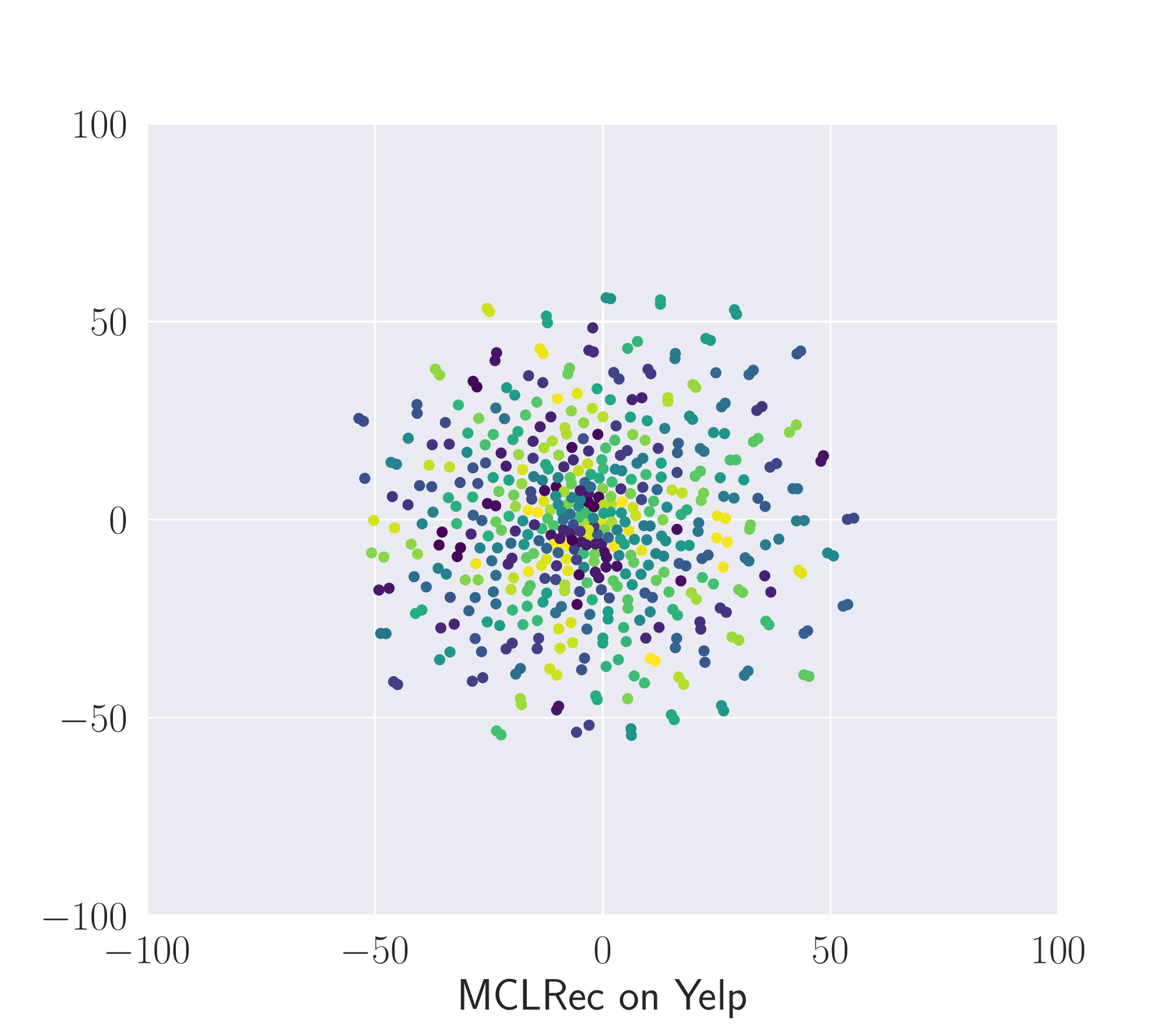}}
    \end{minipage}
       \caption{Comparison of the data augmentation views, $\Tilde{\mathbf{h}}^{1}$ and $\Tilde{\mathbf{h}}^{2}$, trained with different strategies on Sports and Yelp datasets. The dimensions are reduced via T-SNE, where different colors represent negative pairs.}
         \label{meta_exp}
\end{figure}
We use the joint-learning strategy and meta-learning strategy to train our model for 300 epochs in an end-to-end manner respectively and utilize T-SNE to reduce the augmented embeddings into two-dimensional space. 
Limited by the space, the results of Sports and Yelp are presented in Figure~\ref{meta_exp}. 
We intuitively observe that the representations of the negative pairs generated by MCLRec are more "scattered" and the representations of positive pairs generated by MCLRec are "close" than that of MCLRec-J, which indicates that meta-learning strategy helps avoid collapsed results and outputs more informative representations for recommendation. 
The main reason may be that there are two modules (i.e., encoder and augmenters) with parameters that need to be updated, and the target objects of the two modules are different, which leads to a possible gap between the two objects, thus directly using joint learning to update their parameters may lead to suboptimal results by lowering the performance of both modules~\cite{multi}. 
\subsection{Further Analysis (RQ4)}
In this section, we conduct experiments on the Sports and Yelp datasets to verify the robustness of MCLRec. For all models in the following experiments, we only change one variable at a time while keeping other hyper-parameters optimal. 
\subsubsection{\textbf{Impact of Batch Size}.} 
From Figure~\ref{batch_exp}, we can see that reducing the batch size deteriorates the performance of all models.
Comparing SASRec and other models, it can be shown that adding a self-supervised auxiliary task can significantly improve the model's performance with different batch sizes. 
Most importantly, MCLRec's performance with 64 batch size can outperform all other models with 256 batch size on Sports and Yelp.  It indicts that, comparing MCLRec and Cl4SRec, our proposed method can preforms well without of large batch size. 
The reason can be concluded that the introduction of learnable model augmentation allows contrastive learning can be trained with more informative augmentation views including $\Tilde{\mathbf{h}}^{1}$ and $\Tilde{\mathbf{h}}^{2}$,  $\Tilde{\mathbf{z}}^{1}$ and $\Tilde{\mathbf{z}}^{2}$.
\begin{figure}[t]
    \begin{minipage}[h]{0.49\linewidth}
        \centering
        \centerline{\includegraphics[width=1.0\textwidth]{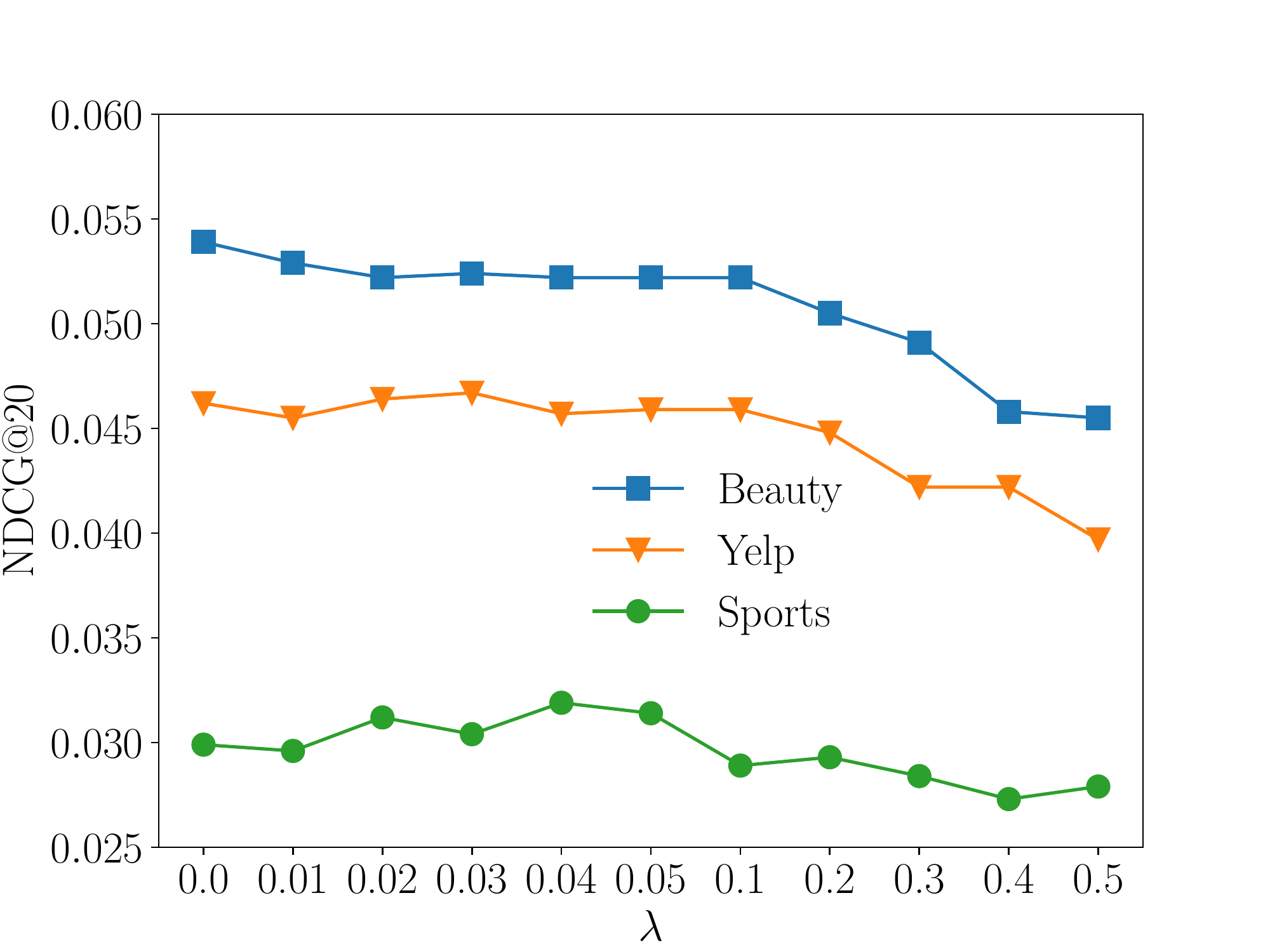}}
    \end{minipage}
        \begin{minipage}[h]{0.49\linewidth}
        \centering
        \centerline{\includegraphics[width=1.0\textwidth]{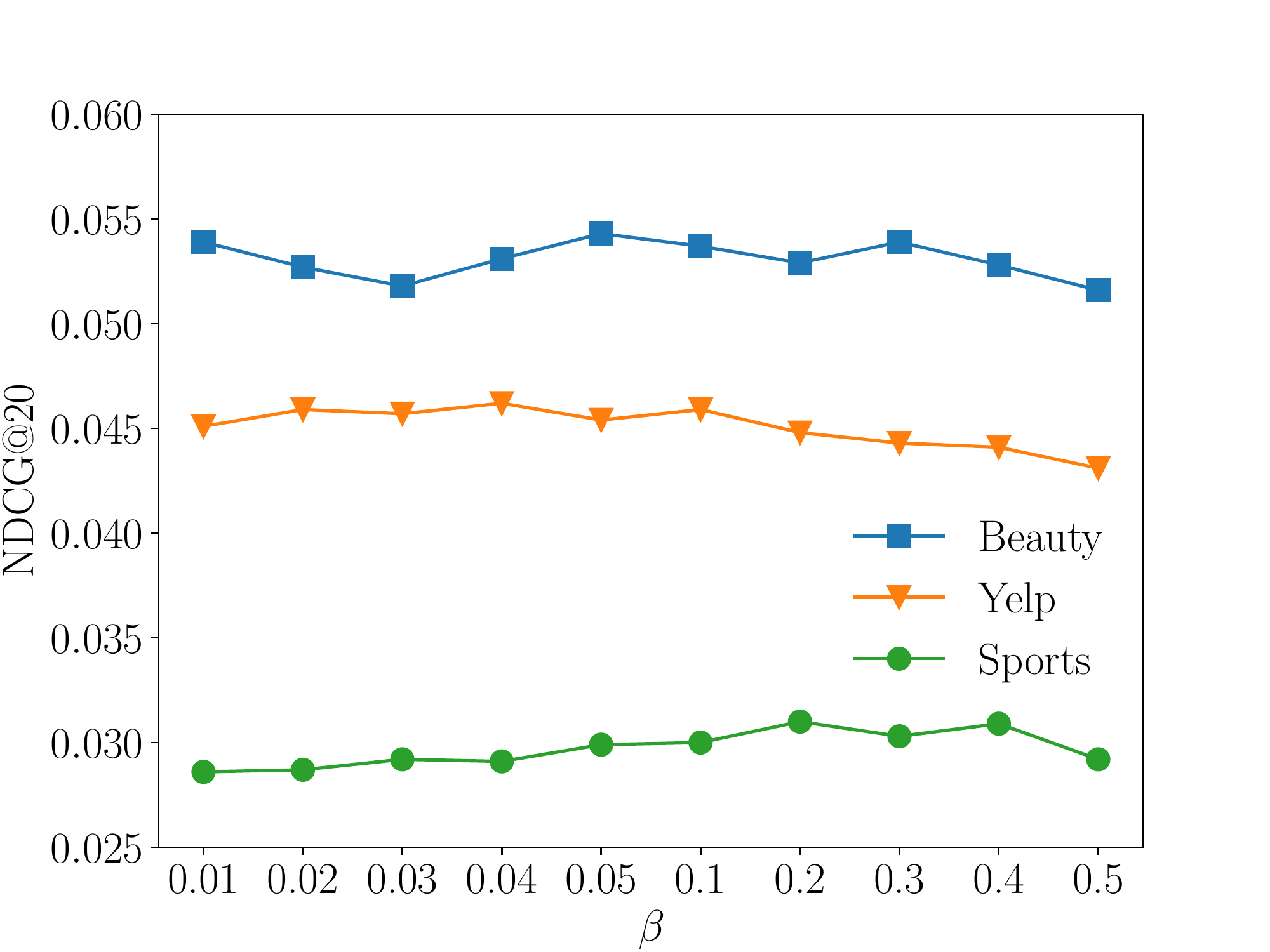}}
    \end{minipage}
      \caption{Performances of MCLRec w.r.t. different weights assigned to the $\mathcal{L}_{cl1}$ and the $\mathcal{L}_{cl2}$ on all datasets.}
         \label{weights_exp}
    \begin{minipage}[h]{0.49\linewidth}
        \centering
    \centerline{\includegraphics[width=1.0\textwidth]{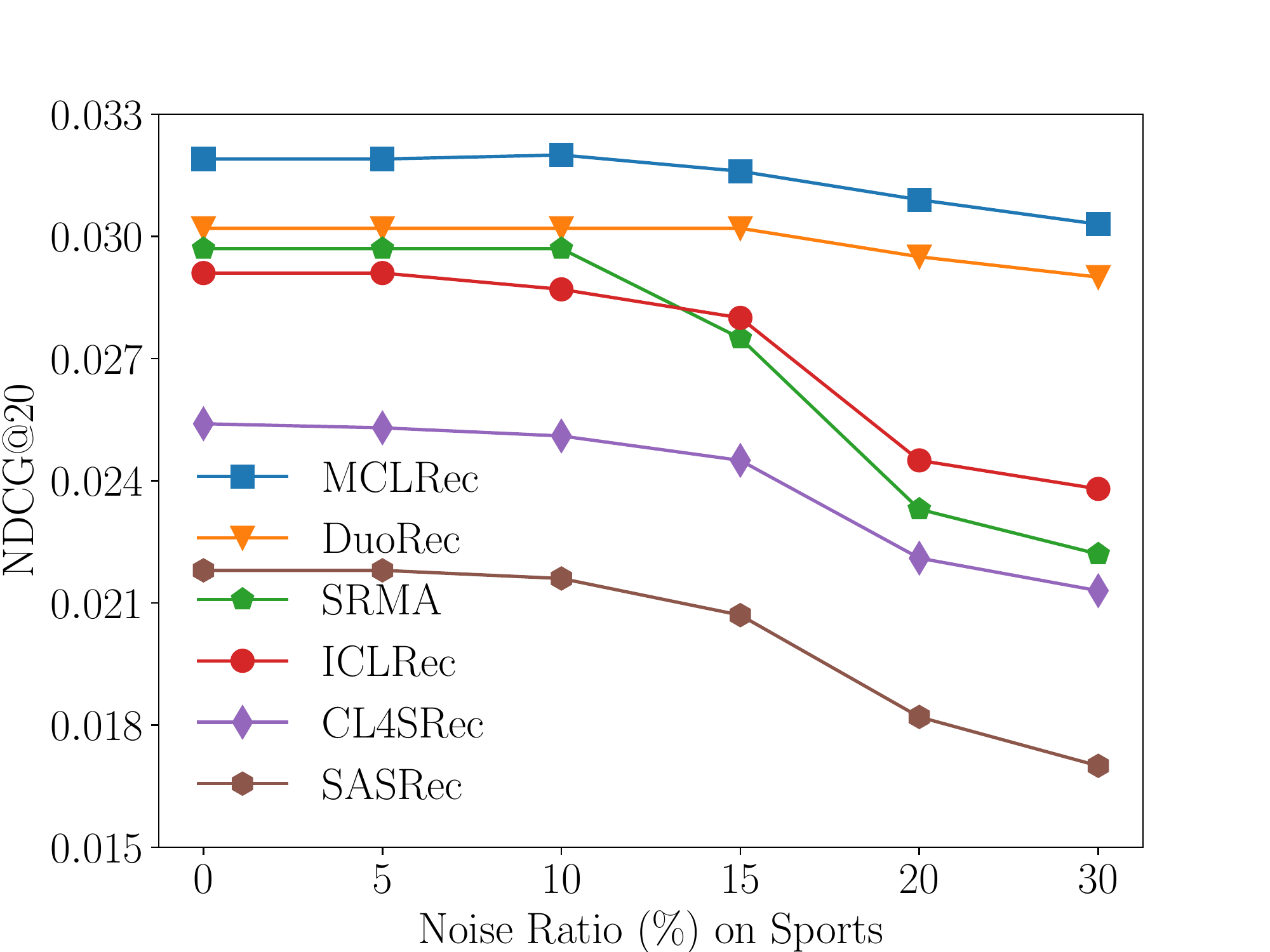}}
    \end{minipage}
    \begin{minipage}[h]{0.49\linewidth}
        \centering
    \centerline{\includegraphics[width=1.0\textwidth]{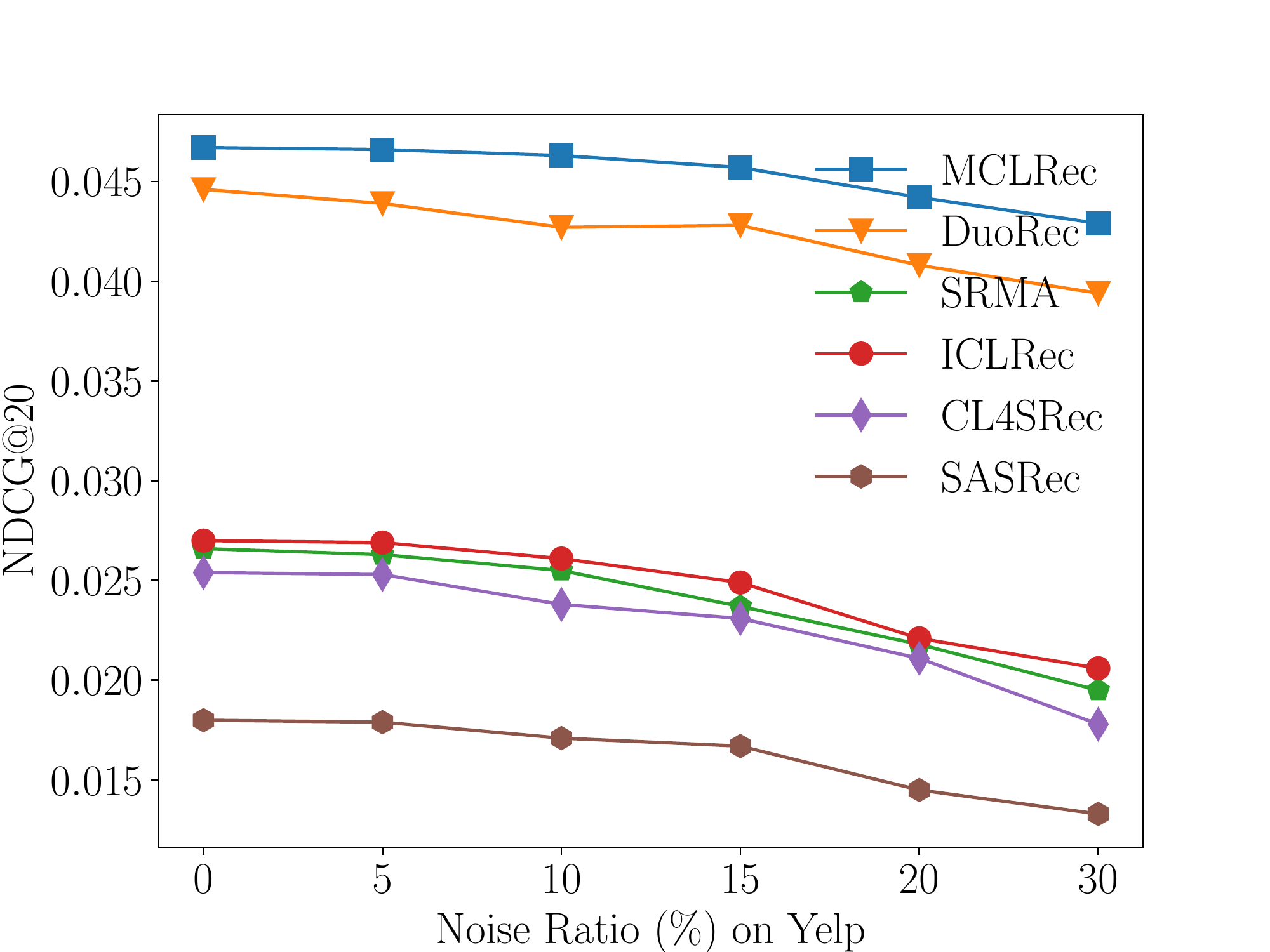}}
    \end{minipage}
      \caption{Performance comparison w.r.t. different Noise Ratio on Sports and Yelp datasets.}
         \label{robustness_exp}
\end{figure}
\subsubsection{\textbf{Hyper-parameters Analysis}.}
The final loss function of MCLRec in Eq.~(\ref{eq13}) is a multi-task learning loss. Figure~\ref{weights_exp} shows the impact of assigning different weights to $\beta$ and $\lambda$ on the model. We observe that the performance of MCLRec gets peak value to different $\beta$ and $\lambda$, which demonstrates the effectiveness of the proposed framework and manifests that introducing suitable weights can boost the performance of recommendation. 
From these figures, $\beta=0.4$ and $\lambda=0.04$ for Sports, $\beta=0.05$ and $\lambda=0$ for Beauty, and $\beta=0.1$ and $\lambda=0.03$ for Yelp are generally proper to MCLRec. The weight of $\mathcal{L}_{cl2}$, i.e., $\beta$, is commonly larger than $\lambda$, which demonstrates that learnable model augmentation generally gains more importance than stochastic data augmentation. 
\subsubsection{\textbf{Robustness to Noise Data}}
To verify the robustness of MCLRec against noise interactions, we randomly add a certain proportion (i.e., $5\%$, $10\%$, $15\%$, $20\%$, $30\%$) of negative items into the input sequences during testing, and examine the final performance of MCLRec and other baselines. 
From Figure~\ref{robustness_exp}, we can see that adding noisy data deteriorates the performance of all models.
By comparing SASRec and other models, it can be seen that adding a 
contrastive self-supervised auxiliary task can significantly improve the model's robustness to noise data. 
By comparing CL4SRec with other models, we can see that introducing model augmentation (e.g., DuoRec, SRMA, and MCLRec) or other auxiliary tasks (e.g., ICLRec) can further alleviate the noise data issues. 
By comparing MCLRec and other models, it can be seen that our model consistently performs better than other models. Especially, with $15\%$ noise proportion, our model can even outperform other models without noise data on two datasets. 
It indicates that, comparing MCLRec with CL4SRec and SRMA, our proposed method can perform well against the noise data. 
The reason can be concluded that with the help of meta training strategy and regular terms, our augmenters can adaptively learn appropriate representations from the stochastic augmented views for contrastive learning. 
\begin{table}[t]
  \centering
    \caption{Statistical information of experimented datasets.}
    \renewcommand{\arraystretch}{1}
  \resizebox{1.0\linewidth}{!}{
    \begin{tabular}{l|c|c|c|c|c|c}
    \hline
    DataSets & \multicolumn{3}{c|}{Sports} & \multicolumn{3}{c}{Yelp} \\
    \hline
    \#length & =5     & 6-8   & >8    & =5     & 6-8   & >8 \\
    \hline
    \#users & 11416 & 14209 & 9973  & 8076  & 11109 & 11246 \\
    \#items & 18357 & 18357 & 18357 & 20032 & 20030 & 20033 \\
    \#actions & 57080 & 95564 & 143693 & 40380 & 75082 & 200892 \\
    sparsity & 99.97\% & 99.96\% & 99.92\% & 99.98\% & 99.97\% & 99.91\% \\
    \hline
    \end{tabular}}%
  \label{tab:sparse}%
\end{table}%
\subsubsection{\textbf{Robustness w.r.t. User Interaction Frequency.}}
\begin{figure}[t]
        \begin{minipage}[h]{0.49\linewidth}
        \centering
        \centerline{\includegraphics[width=1.0\textwidth]{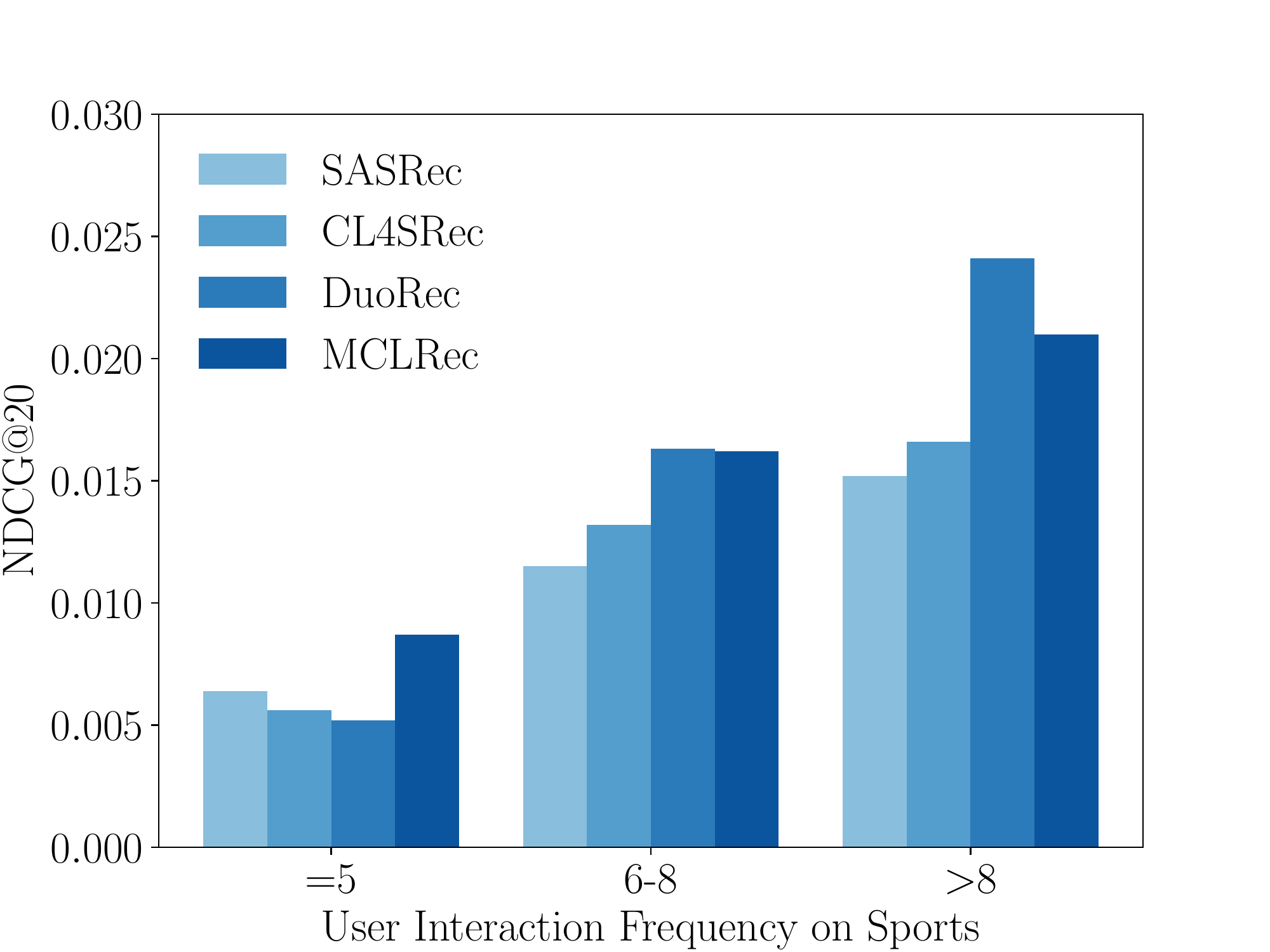}}
    \end{minipage}
    \begin{minipage}[h]{0.49\linewidth}
        \centering
        \centerline{\includegraphics[width=1.0\textwidth]{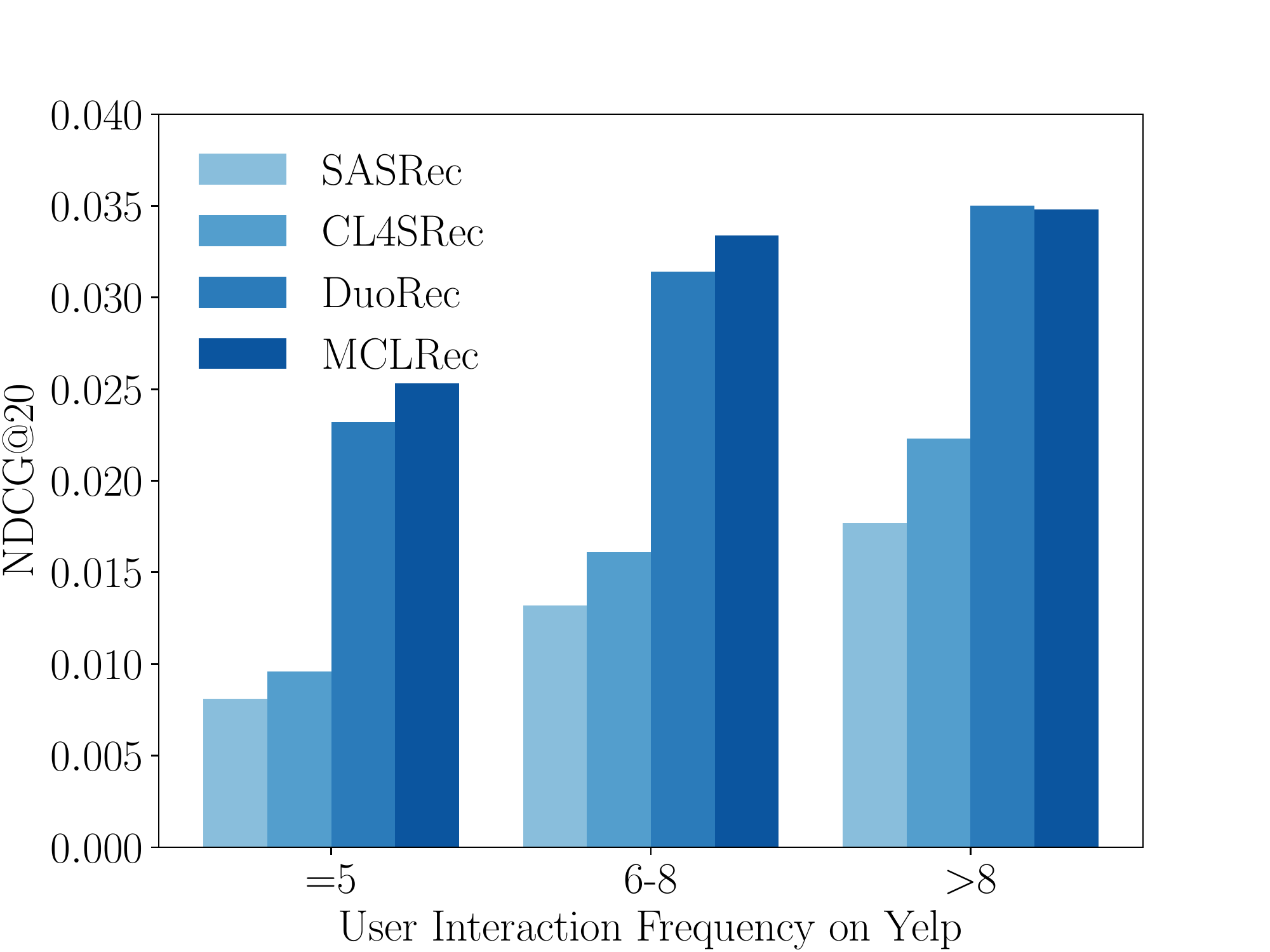}}
    \end{minipage}
      \caption{Performance comparison on different user groups
among SASRec, CL4SRec, DuoRec and MCLRec.}
         \label{sparse_exp}
\end{figure}

To further analyze the robustness of MCLRec against sparse data (e.g., limited historical behaviors), we divide the user behavior sequences into three groups based on their length and keep the total number of behavior sequences constant. The statistics of the prepared datasets are summarized in Table~\ref{tab:sparse}. And all models are trained and evaluated independently on each group of users. From Figure~\ref{sparse_exp}, we observe that reducing the interaction frequency deteriorates the performance of all models. 
By comparing MCLRec with SASRec and CL4SRec, we find that MCLRec can consistently perform better than SASRec and CL4SRec among all user groups. 
This demonstrates that MCLRec can further alleviate the data sparsity problem by introducing more informative augmentation features for contrastive learning, thus consistently benefiting the embedding representation learning even when the historical interactions are limited.
By Comparing MCLRec with the best baseline model DuoRec, it can be seen that the improvement of MCLRec is mainly because it provides better recommendations to users with low interaction frequency. 
This shows that combining data augmentation and learnable model augmentation is beneficial, especially when the recommender system faces the problem of sparse data, where the information of each individual user sequence is limited. 
\section{Related Work}
\subsection{Sequential Recommendation}
Sequential recommendation system~\cite{SRS,SSL} aims to predict successive preferences according to one's historical interactions, which has been heavily researched in academia and industry. Classical Markov Chains~\cite{FPMC}, Recurrent Neural Networks (RNN)-based~\cite{GRU4Rec}, Convolutional Neural Networks (CNN)-based~\cite{Caser}, Transformer\\-based~\cite{SASRec,BERT4Rec,LSAN,STOSA} and Graph Neural Networks (GNN)-based~\cite{SRGNN,SURGE} 
SR models concentrate on users' ordered historical interactions. However, these sequential models are commonly limited by the sparse and noisy problems in practical life. 
\subsection{Self-Supervised Learning for Recommendation}
Motivated by the immense success of Self-Supervised Learning \\(SSL) in Natural Language Process (NLP)~\cite{BERT} and Computer Vision \\(CV)~\cite{VIT,MAE}, and its effectiveness in solving data sparsity problems, a growing number of works are now applying SSL to recommenda-\\tion. 
Among them, some Bidirectional Encoder Representations from Transformer (BERT) like methods to introduce the self-super-\\vised pre-training manner into recommendation~\cite{BERT4Rec,BERT4SessRec}. 
S$^3$-Rec~\cite{S3Rec} introduces four auxiliary self-supervised tasks to capture the sequential information.
Meanwhile, the resurgence of Contrastive Learning (CL) significantly promotes the progress of SSL's research. 
CL4SRec~\cite{CL4SRec} and CoSeRec~\cite{CoSeRec} learns the representations of users by maximizing the agreement between differently augmented views. 
MMInfoRec~\cite{CPC} applies an item level contrastive learning for feature-based sequential recommendation. 
ICLRec~\cite{ICLRec} learns users' intent distributions from users' behavior sequences. 
More prior works also explore the application of CL to graph-based recommendation. 
SGL~\cite{SGL} generates two augmented views with graph augmentation. 
GCA~\cite{GCA} explores adaptive topology-level and node-attribute-level augmentation operations.
DHCN~\cite{DHCN} employs contrastive tasks for hypergraph representation learning.
Different from constructing views only by adopting data augmentation, 
DuoR\\-ec~\cite{DuoRec} chooses to construct view pairs with model augmentation.
LMA4Rec~\cite{LMA4Rec} introduces Learnable Bernoulli Dropout (LBD~\cite{bernoulli}) to the encoder.
SRMA~\cite{SRMA} proposes three levels of model augmentation methods. 
However, these augmentation operations are all hand-crafted and cannot be learned end to end.

\subsection{Meta-Learning for Recommendation}
Meta-learning, which is known as learning to learn~\cite{MAML}, has arouse-d comprehensive interest in recommender systems. 
Most meta-learning-based recommendation models are utilized to initialize the parameters for dealing with the cold-start problems in recommendation systems~\cite{S2meta,CBML,MetaCF,MetaHIN}. 
Recently, some researchers~\cite{Metaselecter,CML,MeLON} have also explored using meta-learning to find optimal hyper-parameters for recommendation. 
For example, 
MeLON~\cite{MeLON} adaptively achieves a better learning rate for new coming user-item interactions. 
Related to meta-learning, our model is designed to update the parameters of learnable augmenters.

\section{Conclusion}
In this paper, we developed a novel contrastive learning-based model called meta-optimized contrastive learning (MCLRec) for sequential recommendation. We took the advantage of data and learnable model augmentation in contrastive learning to create more informative and discriminative features for recommendations. By applying meta-learning, the augmentation model could update its parameters in terms of the encoder's performance. Extensive experimental results showed that the proposed method outperforms the state-of-the-art contrastive learning based sequential recommendation models. In addition, due to the generalization of our framework, in the future, MCLRec could be applied to many other recommendation models and further improve their performance.

\section{ACKNOWLEDGMENTS}
This research was partially supported by the NSFC (61876117, 62176175), the major project of natural science research in Universities of Jiangsu Province (21KJA520004), Suzhou Science and Technology Development Program(SYC2022139), the Priority Academic Program Development of Jiangsu Higher Education Institutions.
\bibliographystyle{ACM-Reference-Format}
\balance
\bibliography{MCLRec}
\end{document}